\documentclass{article}[12pt] 
\usepackage{ amsmath, amsthm, amssymb } 
\usepackage{tikz} 
\usepackage{color} 
\usepackage{cite}

\newcommand{\be} { \begin{equation} } 
\newcommand{\ee} { \end{equation} } 
\newcommand \lless { <{\kern-9pt}< } 
\newcommand{\labbel}[1] { \label{#1} } 
\newcommand{\eps} {\epsilon} 
\newcommand{\D}{\mathcal{D}} 
\newcommand{\ie}{{\it i.e.}} 
\newcommand{\atan} {\mbox{\rm atan}} 
\begin{document} 
\setlength{\unitlength}{1.3cm} 
\begin{titlepage}
\vspace*{-1cm}
\vskip 3.5cm
\begin{center}
\boldmath
 
{\Large\bf The maxcut of the sunrise with different masses in the 
continuous Minkoskean dimensional regularisation} 
\unboldmath
\vskip 1.cm
{\large Filippo Caleca}$^{a,b}$ 
\footnote{{\tt e-mail: filippo.caleca@gmail.com}} 
{\large and Ettore Remiddi}$^{a,}$
\footnote{{\tt e-mail: ettore.remiddi@bo.infn.it}}
\vskip .7cm
{\it $^a$ DIFA, Universit\`a di Bologna and INFN, Sezione di Bologna, 
	I-40126 Bologna, Italy } \\
{\it $^b$ Univ Lyon, Ens de Lyon, CNRS, Laboratoire de Physique, F-69342 Lyon, France } \\
\end{center}
\vskip 2.6cm

\begin{abstract}
We evaluate the {\it maxcut} of the two loops sunrise amplitude with
three different masses by using the {\it Minkoskean} (as opposed to
the usual {\it Euclidean}) continuous dimension regularisation,
obtaining in that way six related but different functions expressed in
the form of onedimensional finite integrals.
We then consider the $4$th order homogeneous equation valid for the maxcut, 
and show that for arbitrary dimension $d$ the six functions do satisfy the 
equation. We further discuss the $d=2,3,4$ cases, verifying that only four 
of them are linearly independent.
The equal mass limit is also shortly discussed.  

\end{abstract} 
\vfill 
\end{titlepage} 
\newpage 
\section{Introduction} \labbel{sec:intro} \setcounter{equation}{0} 
\numberwithin{equation}{section} 
As known, one can derive an inhomogeneous differential equation for any 
Feynman amplitude. The solution of that equation can then be obtained by the 
Euler's variation of the constants method, which relies on the 
knowledge of the solutions of the associated homogeneous equation. 
The so-called {\it maxcut} amplitude ({it i.e.} the loop integral in which 
all the propagators are replaced by the corresponding $\delta$-functions) 
provides always with a solution of the homogeneous equation. 
(For the introduction and use of the {\it maxcut}, 
see~\cite{Primo:2017ipr} ). 
\par 
But it can happen, within the usual continuous dimensional regularisation 
approah, that the information so obtained is not sufficient (if the 
differential equation is of higher order) to find all the solutions, 
or even that the evaluation of the {\it maxcut} gives a vanishing result, 
because the constraints of the $\delta$-functions lead to a vanishing 
integration range for the loop variables. 
\par 
To overcome the problem, one can give~\cite{Remiddi2022} a {\it Minkowskean} 
metric to the regularising components, as opposed to the {\it Euclidean} 
metric which id always implicitly used. (Let us recall, in this context, 
that the regularising loop components do not mix with the components of the 
physical external momenta, so that Wick rotations can be safely carried 
out). It is found that, when the {\it Minkowskean} metric is used, the 
evaluation of the {\it maxcut} gives a full information on all the solutions 
of the homogeneous differential equation. 
\par 
We consider in this paper the 
case of the sunrise with three different masses, and we show that the 
{\it maxcut} evaluated within the {\it Minkowskean} metric approach 
does in fact provide with a full information on all the solutions of 
the 4th order homogeneous equation. 
Some preliminary work~\cite{Remiddi2022}  indicates that the new 
{\it Minkowskean} approach could apply, more in general, to three and four 
point amplitudes {\it etc.} as well. 
\par 
In the considered case of the sunrise with three different masses 
and in $d$-continuous dimensions, the 
proposed {\it Minkowskean} approach evaluates the {\it maxcut}, 
for each value of the external physical momentum, as the sum 
of 6 functions. We verify that each of them separately satisfies 
the homogeneous 4th order equation equation, and for $ d =2,3,4 $ we 
verify also that four of them are linearly independent. 
\par 
The mathematical tools which we use are almost elementary (such as integration 
by parts identities), but many of the formulas are in general too cumbersome 
to be reported explcitly. Correspondingly, for the sake of clarity, our 
treatment will be occasionally somewhat pedagogical. 
\par 
The plan of the paper is as follows. \\ 
Section 2 recalls the 4th order equation for the {\it maxcut} of the 
Sunrise amplitude. \\ 
Section 3 describes the evaluation of the {\it maxcut}. \\ 
Section 4 verifies that the functions introduced in the previous Section 
satisfy the 4th order homogeneous equation. \\ 
Section 5 deals with the $ d = 2 $ case. \\ 
Section 6 deals with the $ d = 3 $ case. \\ 
Section 7 deals with the $ d = 4 $ case. \\ 
Section 8 discusses the equal mass Sunrise amplitude. \\ 
Appendix A describes the explicit evaluation of the {\it maxcut}. \\ 
Appendix B gives the explicit expression of the 4th order homogeneous 
equation. \\ 
Appendix C discusses the relations between the functions of the $ d = 2 $ 
case with the elliptic  functions. 
\section{The Equations for the Sunrise amplitude and its Maxcut} 
\labbel{sec:Sun} \setcounter{equation}{0} 
\numberwithin{equation}{section} 
Let us write the Sunrise aplitude with different masses in the continuous 
dimensional regularisation as 
\be S(d,s) = \int \D^dk\int \D^dq\ 
         \frac{1}{D_1-i\eps}\ \frac{1}{D_2-i\eps}\ \frac{1}{D_3-i\eps} \ , 
\labbel{eq:S} \ee 
where: \\ 
$k$ and $q$ are the integration loop momenta; 
\begin{align} 
   D_1 &= q^2 + m_1^2 \ , \nonumber\\ 
   D_2 &= (k-q) + m_2^2 \ , \nonumber\\ 
   D_3 &= (p-k)^2 + m_3^2 \ ; \labbel{eq:D123} 
\end{align} 
$ (m_1, m_2, m_3 ) $ are the three (in principle different) masses 
of the three propagators; \\  
$ p $ is the external momentum, and 
\be s = - p^2 = p_0^2 -\vec{p}^2 \ . \labbel{eq:s} \ee 
\par 
Well established methods and computer programs are available for deriving a 
differential equation in $ s $ for the amplitude $ S(d,s) $; by using 
LiteRed~\cite{Lee2012} we obtained a 4th order inhomogneous linear 
differential equation for $ S(d,s) $, which reads 
\be D_4(d,s)\ S(d,s) = T(d,s) \ , \labbel{eq:eqS}    \ee 
where $ T(d,s) $ is the imhogeneous term (a sum of tedpoles) which will be 
ignored in the rest of this paper, and $ D_4(d,s) $ is 
a 4th order differential operator which can be written as 
\begin{align} 
      D_4(d,s) &=  C_4(d,s) \frac{d^4}{ds^4} 
                 + C_3(d,s) \frac{d^3}{ds^3} \nonumber\\ 
                &+ C_2(d,s) \frac{d^2}{ds^2} 
                 + C_1(d,s) \frac{d}{ds} 
                 + C_0(d,s)\ , \labbel{eq:D4} 
\end{align} 
the 5 coefficients $ C_i(d,s6), i=0,1,..,4 $ being 
polynomials in $d, s$ and in the three masses $m_k, k=1,2,3$. 
The explicit expressions of those polynomials, which are rather 
cumbersome, are explicitly written in Appendix \ref{app:equ}. 
\par 
Let us remind here that the Euler's variation of constants method, commonly 
used for the solution of an inhomogeneous 4th order linear differential 
equation as Eq.(\ref{eq:eqS}), requires the knowledge of 4 linearly 
independent solutions of the associated 4th order homogeneous equation, 
which in our case is 
\be D_4(d,s)\ \phi(d,s) = 0 \ . \labbel{eq:ahe} \ee 
\par 
We consider now the maxcut of the Sunrise amplitude, which in the notation 
of Eq.(\ref{eq:S}) is 
\be M(d,s) = \int \D^dk\int \D^dq\ \delta(D_1)\delta(D_2)\delta(D_3) \ . 
\labbel{eq:M} \ee 
By repeating the procedure already used for obtaining the inhomogeneous 
differential equation for the Sunrise amplitude, one finds that the 
maxcut $ M(d,s) $ satisfies the above homogeneous equation 
\be D_4(d,s) \ M(d,s) = 0 \ . \labbel{eq:D4M} \ee 
Indeed, the inhomogeneous part of the equation for $ S(d,s) $ is due to the 
presence, in the integration by parts identities used in the derivation of 
the inhomogeneous equation, of terms of the form 
\be   D_i \times \frac{1}{D_i-i\eps} = 1 \ ;    \labbel{eq:DiP} \ee 
the same derivation (with the same integration by parts identities) 
can be used for deriving the equation for the maxcut $ M(d,s) $, in which 
all the propagators $ 1/(D_i-i\eps) $ are replaced by the corresponding 
$\delta$-function $ \delta(D_i) $, so that Eq.(\ref{eq:DiP}) becomes 
\be   D_i \times \delta(D_i-i\eps) = 0 \ ,  \labbel{eq:DiD} \ee 
and inhomogeneous terms drop out. 
\par 
We will show in the next Section how to evaluate the maxcut $ M(d,s) $, 
and in the subsequent Sections we will discuss the solutions of 
Eq.(\ref{eq:D4M}) and their properties. 
\section{The Maxcut} 
\labbel{sec:Mc} \setcounter{equation}{0} 
\numberwithin{equation}{section} 
We now evaluate the Maxcut $ M(d,s) $ defined by Eq.(\ref{eq:M}), with 
$ D_1, D_2, D_3 $ given by Eq.s(\ref{eq:D123}). \par 
We write the components of the 4 dimensional vector $p$, in full generality, 
as 
\be p = (p_0, p_x,0,0) , \labbel{eq:pcomp} \ee 
\ie setting $ p_y = p_z = 0 $. As $ p $ does not span the $y$ and $z$ 
directions, we write 
the components of the continuous $d$-dimensional loop integration momenta 
as 
\begin{align} 
    k &= (k_0,k_x,K_n), {\hskip5mm} n = 1,2,..,d-2 \ , \nonumber\\ 
    q &= (q_0,q_x,Q_n), {\hskip5mm} n = 1,2,..,d-2 \ , \labbel{eq:kq} 
\end{align} 
where we have merged the $ y $ and $ z $ components of $ k $ and $ q $ 
and their $ (n-4) $ regularising components into the $(d-2)$-dimensional 
vectors $(K_n,Q_n)$, so that one has 
\be 
    \D^dk = dk_0\ dk_x\ d^{d-2}K \ , \labbel{eq:Ddk} 
\ee 
\be 
    \D^dq = dq_0\ dq_x\ d^{d-2}Q \ . \labbel{eq:Ddq} \ee 
In the usual dimensional regularisation, the regularising components 
are assumed to have an Euclidean metric; in this paper, however, 
we give to $(\vec{K},\vec{Q})$ a Minkowskean metric, so that 
\begin{align} 
     k^2 &= -k_0^2 +k_x^2 - K^2 \ , \nonumber\\ 
     q^2 &= -q_0^2 +q_x^2 - Q^2 \ , \nonumber\\ 
     (k\cdot q) &= - k_0q_0 + k_xq_x - (\vec{K}\cdot \vec{Q}) \ . 
                                                           \labbel{eq:k2} 
\end{align} 
A few comments on that choice. 
As the external physical vector $p$ spans only the time and $x$ 
dimensions, scalar products of $p$ and the components of the 
regularising vectors $(\vec{K},\vec{Q})$ never occur, and, 
quite in general, the metric of the components of the loop momenta 
corresponding to the dimensions not spanned by the physical vectors 
does not matter and can be changed at will from 
Euclidean to Minkowskean (Wick rotation). 
\par 
As a further comment, let us recall that a propagator 
is equal to a Principal value plus a $\delta$-function 
\be \frac{1}{D-i\eps} = \frac{\mathcal{P}}{D} + i\pi\delta(D) 
\labbel{eq:ProPD} \ . \ee 
In the amplitude with full propagators $ 1/(D-i\eps) $, as 
$ S(d,s) $ of Eq.(\ref{eq:S}), the above change of the metric does not affect 
the value of the amplitude, but the change does affect the maxcut 
Eq.(\ref{eq:M}), which involves only the $\delta$-functions 
(and not the complete propagators). 
\par 
At any rate, in the rest of this paper we will be concerned only with the 
maxcut, defined by Eq.(\ref{eq:M}) and evaluated with regularising 
loop momenta with Minkoskean metric, independently of possible Wick rotations 
of the complete amplitude Eq.(\ref{eq:S}). We will therefore limit ourselves 
to the study of the solutions of Eq.(\ref{eq:D4M}), where 
however $ D_4(d,s,m_i^2) $ is exactly the differential operator appearing 
in the inhomogeneous differential equation valid for the amplitude 
Eq.(\ref{eq:eqS}). 
\par 
After those preliminary considerations, let us start the actual evaluation 
of the maxcut by defining 
\be B(d,-k^2) = \int \D^dq\ \delta(D_1)\delta(D_2) \ . \labbel{eq:Bk2} \ee 
Eq.(\ref{eq:M}) then becomes 
\be M(d,s) = \int \D^dk \ B(d,-k^2) \ \delta(D_3) \ ; \nonumber \ee 
by further introducing a factor 
\be \int db \ \delta(b+k^2) = 1 \ , \nonumber \ee 
(it is understood that the integration in $b$ ranges from $-\infty$ 
to $+\infty$) and suitably interchanging the orders of integration one has 
\begin{align} 
  M(d,s) &= \int \D^dk\int db \ B(d,-k^2)\ \delta(b+k^2) \ \delta(D_3) 
                                                   \nonumber\\ 
  &= \int db \ B(d,b)\ C(d,-p^2,b) \ , \labbel{eq:Mb} 
\end{align} 
where we have introduced 
\be C(d,-p^2,b) = \int \D^dk\ \delta(b+k^2) \ \delta(D_3) \ . 
\labbel{eq:C} \ee 
\par 
For simplicity, in this paper we discuss only the case of 
timelike $p$ above threshold, \ie with 
\be - p^2 = s = E^2, {\hskip5mm} E>(m_1+m_2+m_3). \labbel{eq:Em1m2m3} \ee 
Referrinng to Appendix \ref{app:max} for more details of the integrations 
(which are essentially elementary), 
we report here the explicit expression of $ M(d,s) $ as an integral in $b$, 
in the case of different masses, ordered as 
\be  m_1 \ge m_2 \ge m_3. \labbel{eq:mi}       \ee 
To that aim, and for the subsequent discussions of the paper, we 
introduce the 6 constants (with respect to $b$) 
\begin{align} b_{-1} &= -\infty     \ ,\nonumber\\ 
                 b_0 &= 0           \ ,\nonumber\\ 
                 b_1 &= (m_1-m_2)^2 \ ,\nonumber\\
                 b_2 &= (m_1+m_2)^2 \ ,\nonumber\\ 
                 b_3 &= (E-m_3)^2   \ ,\nonumber\\ 
                 b_4 &= (E+m_3)^2   \ ,\nonumber\\ 
                 b_5 &= +\infty     \ ,\labbel{eq:bi} 
\end{align} 
which for $ E $ above threshold, Eq.(\ref{eq:Em1m2m3}), are in 
increasing order 
\be b_{-1} < b_0 < b_1 < b_2 < b_3 < b_4 < b_5 \ , \labbel{eq:biord} \ee 
and the the 4th order polynomial 
\be R_4(b) = (b-b_1)(b-b_2)(b-b_3)(b-b_4) \ . \labbel{eq:R4} \ee 
The result of our calculation of $ M(d,s) $ can then be written 
as the sum of 6 terms 
\be M(d,s) = \sum_{i=0,5} C_i(d)\ M_i(d,s) \ , \labbel{eq:Mc} \ee 
where: \\ 
the explicit expressions of the factors $ C_i(d) $, which depend only on $d$ 
(albeit not needed for the discussions which will follow) are given in 
Eq.s(\ref{eq:defCi}) ; \\ 
the $ M_i(d,s) $ are the following 6 quantities, involving 
6 one dimensional integrals 
\be  M_i(d,s) = s^\frac{2-d}{2} \int_{b_{i-1}}^{b_i} db 
            \ |b|^\frac{2-d}{2}\ |R_4(b)|^\frac{d-3}{2} \ . 
\labbel{eq:Mi} \ee 
The 6 integrals span the whole integration range of the integration 
variable $b$ from $-\infty$ to $+\infty$. The convergence of the separate 
integrals depends on the behaviour of the integrands at the end points 
of integration, which in turn depend on $d$. In particular: \\ 
the convergence of $ M_0(d,s) $, whose integration range is 
$ -\infty < b < 0 $, requires $ d < 8/3 $ at $ b = b_{-1} = -\infty $ and 
$ d < 4 $ at $ b = b_0 = 0 $ ; \\ 
$ M_1(d,s) $ requires $ d < 4 $ at $ b_0 = 0 $ and $ d > 1 $ 
                                              at $ b = b_1 $; \\ 
$ M_2(d,s), M_3(d,s) $ and $ M_4(d,s) $ require $ d > 1 $; \\ 
$ M_5(d,s) $ (range $ b_4 < b < b_5 = +\infty $) 
requires $ d > 1 $ at $ b = b_4 $ and $ d< 8/3 $ at $ b = b_{5} $; \\ 
Therefore all the 6 integrals and their sum $ M(d,s) $, Eq.(\ref{eq:Mc}), 
converge for $ d $ within the range $ 1 < d < 8/3 $. 
The value of $ M(d,s) $ as given by Eq.(\ref{eq:Mc}), is therefore 
well defined, and the maxcut $ M(d,s) $ of Eq.(\ref{eq:M}) 
is a solution of the 4th order homogenous equation Eq.(\ref{eq:D4M}).  
\par 
Let us recall here that the imaginary part of $ S(d,s) $ (or, more precisely, 
its discontinuity in $ s $) in the usual Euclidean regularisation scheme 
is equal, up to a constant ($s$ independent) factor, to 
\begin{align} 
   {\rm Im} S(d,s) &= \theta\left(+E-(m_1+m_2+m_3)\right) 
                      \times \int \D^dk\int \D^dq\ 
                    \ \theta(+q_0)\delta(D_1) 
                    \ \theta(k_0-q_0)\delta(D_2) 
                    \ \theta(p_0-q_0)\delta(D_3)         \nonumber\\ 
                   &+ \theta\left(-E-(m_1+m_2+m_3)\right) 
                      \times \int \D^dk\int \D^dq\ 
                    \ \theta(-q_0)\delta(D_1) 
                    \ \theta(q_0-k_0)\delta(D_2) 
                    \ \theta(q_0-p_0)\delta(D_3) \ , \labbel{eq:ImS} 
\end{align} 
which, when evaluated, is found to be proportional to the single integral 
$ M_3(d,s) $ defined above, see Eq.(\ref{eq:Mi}), whose integration range 
in $ b $ is $ (m_1+m_2)^2 < b < (E-m_3)^2 \ . $ 
\par 
On the other hand, $ {\rm Im} S(d,s) $ is a solution of the homogeneous 
equation Eq.(\ref{eq:ahe}), as the inhomogeneous term $ T(d,s) $ of the 
complete equation Eq.(\ref{eq:eqS}) has vanishing discontinuity. 
Therefore, $ M_3(d,s) $ alone, independenty of the other $ M_i(d,s) $, 
is also a solution of the homogeneous equation. 
\par 
Further, again in the conventional Euclidean regularization, (at variance 
with the Minkowskean regularisation used in this paper) one finds (we 
omit details for obvious space problems) that the whole maxcut 
is given by a linear combination, similar to Eq.(\ref{eq:Mc}), 
of the 4 integrals $ M_0(d,s) $, $ M_1(d,s) $, $ M_3(d,s) $ 
and $ M_5(d,s) $ only~\cite{Caleca2020}. 
\par 
One is then naturally led to guess that any of the 6 terms $ M_i(d,s), $ 
separately taken, is a solution of that 4th order equation; on the other hand, 
it is known that a 4th order equation can have only 4 independent solutions, 
so that the 6 terms cannot be all linearly independent. 
Those issues will be discussed in the rest of this paper. 
\section{Derivatives and differential equations of the $b$ integrals 
for arbitrary $d$} 
\labbel{sec:DDd} 
\setcounter{equation}{0} 
\numberwithin{equation}{section} 
In this section we show how to evaluate the 4 $s$-derivatives 
of each of the 6 functions $M_i(d,s)$, (\ref{eq:Mi}), 
and check that each of the 6 functions does separately satisfy the 4th order 
differential equation Eq.(\ref{eq:ahe}). 
The 6 equations for each of the 6 $ M_i(d,s) $ 
can be seen as an identity involving 5 functions, namely $M_i(d,s)$ 
and its first 4 derivatives. 
By means of suitable integration by parts identities in $ b $, 
we will then express those 5 functions in terms of a set of 4 auxiliary 
functions only, we will substitute those expressions in the 
4th order equation, which becomes a combination of those 4 auxiliary 
functions, and then we will verify that after the substitutions 
each of the six 4th order equations does in fact vanish. 
\par 
To marginally simplify notation and formulas, we introduce the 6 new 
functions, for $ i=0,1,..,5 $, 
\be  N_i(d,s) = \int_{b_{i-1}}^{b_i} db 
            \ |b|^\frac{2-d}{2}\ |R_4(b)|^\frac{d-3}{2} \ , 
\labbel{eq:Ni} \ee 
so that, obviously, Eq.(\ref{eq:Mi}) becomes 
\be M_i(d,s) = s^{\frac{2-d}{2}} N_i(d,s) \ , \labbel{eq:MNi} \ee 
and we will discuss the derivatives of the $ N_i(d,s) $ instead of the 
derivatives of $ M_i(d,s) $. \par 
As the $b$ integral $ N_i(d,s) $ depend on the 7 parameters $ b_j $, 
$ j=-1,0,..,5 $, and therefore depend on $s = E^2 $ through the presence of $s$ 
in $ b_3, b_4 $, see Eq.s(\ref{eq:bi}), we have quite in general 
\be \frac{d}{ds}N_i(d,s) = \left( \frac{db_3}{ds}\frac{d}{db_3} 
       + \frac{db_4}{ds}\frac{d}{db_4} \right) N_i(d,s) \ ; 
\labbel{eq:dsNi} \ee 
we will therefore look for derivatives of the integrals $ N_i(d,s) $ 
with respect to $ b_3 $ and $b_4$. The approach could be easily extended 
also to $b_1$ and $b_2$, allowing the evaluation of the derivatives with 
respect to the masses, which however we will not consider in this paper. 
\par 
We discuss first the derivatives of the 3 integrals 
\be N_n(d,s), \hskip1cm n=2,3,4 \labbel{eq:Nn} \ee 
whose integration ranges are $ b_1 < b < b_2 $, $ b_2 < b < b_3 $ and 
$ b_3 < b < b_4 $. \par 
If $ b_3 $ or $ b_4 $ is an end-point of the integration, a first 
contribution to the derivatives of those integrals can arise, in principle, 
from the corresponding end-point value of the integrands. 
As already observed in Section \ref{sec:Mc}, the 3 integrals listed in 
Eq.(\ref{eq:Nn}) converge for $ d > 1 $; 
therefore, we can consider the $ d > 3 $ case, so that the contributions from 
the end-point values of the derivatives in $b_3$ and $b_4$ of those integrals 
vanish (and by analitic continuation we can extend the final results also 
to $ 1 < d \le 3 $). 
We then obtain for $ k=3,4 $  
\be \frac{d}{db_k} N_n(d,s) = \int_{b_{n-1}}^{b_n} db 
            \ |b|^\frac{2-d}{2}\ |R_4(b)|^\frac{d-3}{2} 
            \times \left( - \frac{1}{b-b_k} \right) \ , \labbel{eq:bkNn} \ee 
which converges for $ d > 3 $. But we need also higher order derivatives, 
up to the 4th; by simply taking repeatedly the $ d/db_k $ derivatives of 
Eq.(\ref{eq:bkNn}) (with higher vales of $d$, for convergence sake), 
we would get up to 4 powers of the denominator $ 1/(b-b_k) $, which 
including $ N_n(d,s) $ gives a total of 5 apparently unrelated 
functions, while our purpose is to express the $ N_n(d,s) $ and their 
4 derivatives in terms of 4 independent functions only. 
\par 
To overcome that problem, for each of the 3 integrals $ N_n(d,s) $ 
of Eq.(\ref{eq:Nn})  we consider a set of 4 {\it auxiliary integrals} 
\be N_n^{(j)}(d,s) = \int_{b_{n-1}}^{b_n} db 
            \ |b|^\frac{2-d}{2}\ |R_4(b)|^\frac{d-3}{2} 
            \times b^j \ , \labbel{eq:Nnj} \ , \ee 
where $ j = -1,0,1,2 $, and $N_n^{(0)}(d,s) = N_n(d,s) $. \par 
We then write the following 4 {\it integration by part identities} in $ b $ 
\be \int_{b_{n-1}}^{b_n} db \left( \delta(b-b_n)-\delta(b-b_{n-1}) 
                                 - \frac{d}{db}\right) 
    \ |b|^{\frac{2-d}{2}}\ |R_4(b)|^\frac{d-3}{2} 
            \times \left(  \frac{|R_4(b)|}{b-b_k} \right) = 0 \ . 
\labbel{eq:bnbp} \ee 
For $ d > 1 $ all the end-point contributions 
vanish; considering for definiteness the case $ b_k = b_3 $ and 
by evaluating explicitly the $b$ derivative, 
we can derive from Eq.(\ref{eq:bnbp}), after some algebra, the relation 
\begin{align} 
 \int_{b_{n-1}}^{b_n} db& \ |b|^\frac{2-d}{2}\ |R_4(b)|^\frac{d-3}{2} 
 \times \frac{1}{b-b_3} = \frac{2}{(d-3)(b_3-b_1)(b_3-b_2)(b_4-b_3)} 
 \int_{b_{n-1}}^{b_n} db \ |b|^\frac{2-d}{2}\ |R_4(b)|^\frac{d-3}{2} 
\nonumber\\ 
 \times \Biggl[& + \left(\frac{3}{2}d-2\right)\ \times b^2 \nonumber\\ 
 & + \left( \frac{d-3}{2}b_3 - \left(d-\frac{3}{2} \right)(b_1+b_2+b_4) 
     \right) \times b 
\nonumber\\ 
 & + \left( - \frac{d-3}{2} b_3 (b_1+b_2-b_3+b_4) 
            + \frac{d-2}{2} (b_1b_2 + b_1b_4 + b_2b_4) \right) 
             \times 1 \nonumber\\ 
 & + \frac{d-2}{2}b_1b_2b_4 \times \frac{1}{b} \Biggr] \ , \labbel{eq:bmb3} 
\end{align} 
which can be used to rewrite the $ b_3 $ derivative of each of the 
3 functions $ N_n(d,s) = N_n^{(0)}(d,s), n=2,3,4 $, see Eq.(\ref{eq:bkNn}) for 
$ b_k = b_3 $, in terms of the 4 {\it auxiliary integrals} 
$ N_n^{(j)}(d,s),\ j=-1,0,1,2,$ Eq.(\ref{eq:Nnj}). 
\par 
Similar formulas can also be easily derived from Eq.(\ref{eq:bnbp}) 
for the other values of 
$ b_k $, namely $ (b_1,b_2,b_4) $, and can be used for the derivatives of the 
$ N_n^{(0)}(d,s) $ with respect to those values of $b_k$. It is to be 
noted that, although those formulas for the derivatives, in which the 
$ 1/(b-b_k) $ factors are no longer present, were established for $ d > 3 $, 
they are are actually valid also for $ d > 1 $. 
\par 
But the same happens also for all the $ b_k $ derivatives of all the 
other $ N_n^{(j)}(d,s) $, as it is immediate to check. Indeed, one has, 
for instance 
\begin{align} 
  \frac{d}{db_k} N_n^{(1)}(d,s) &= \int_{b_{n-1}}^{b_n} db 
    \ |b|^\frac{2-d}{2}\ |R_4(b)|^\frac{d-3}{2} 
    \times \left( - 1 - \frac{b_k}{b-b_k} \right) \ , \nonumber\\ 
  \frac{d}{db_k} N_n^{(-1)}(d,s) &= \int_{b_{n-1}}^{b_n} db 
    \ |b|^\frac{2-d}{2}\ |R_4(b)|^\frac{d-3}{2} 
    \frac{1}{b_k}\times \left( \frac{1}{b} - \frac{1}{b-b_k} \right) \ , 
\nonumber  
\end{align} 
with a similar formula holding for $ N_n^{(2)}(d,s) $ as well. 
\par 
Summarising, by replacing systematically the integrals containing the factor 
$ 1/(b-b_k) $ with the proper analog of Eq.(\ref{eq:bmb3}), all the $b_k$ 
derivatives of all the integrals $ N_n^{(j)}(d,s) $ are then completely 
expressed in terms of the same $ N_n^{(j)}(d,s) $ only. 
By suitably iterating the procedure, one can express all the higher order 
derivatives of the 3 functions $ N_n(d,s) $ of Eq.(\ref{eq:Nn}) with respect 
to all the 4 parameters $b_k$ in terms of the 4 $ N_n^{(j)}(d,s) $ 
only. 
\par  
By recalling Eq.s(\ref{eq:MNi}),(\ref{eq:dsNi}) on can then obtain all the 
$s$ derivatives of the $ M_i(d,s) $, Eq.(\ref{eq:Mi}), for $ i=2,3,4 $, 
in terms of the corresponding 4 {\it auxiliary integrals} Eq.(\ref{eq:Nnj}). 
The explicit expressions are rather clumsy, and involve too many terms 
(from 189 of the 1st $s$ derivative up to almost 17 thousand of the 
4th) to be written here; but by using those expressions one can 
check that indeed the 3 integrals $ M_2(d,s), M_3(d,s), M_4(d,s) $ do 
satisfy, separately, the 4th order differential equation Eq.(\ref{eq:ahe}). 
\par 
We consider next the derivatives with respect to $b_3$ and $b_4$ of 
\be  N_1(d,s) = \int_{b_0=0}^{b_1} db 
                \ |b|^\frac{2-d}{2}\ |R_4(b)|^\frac{d-3}{2} \ , 
\labbel{eq:N1} \ee 
which due to the end-points $(0,b_1) $ converges in the range $ 1 < d < 2 $. 
For this case we take as {\it auxiliary integrals}, instead of the 
Eq.s(\ref{eq:Nnj}), the 4 integrals, for $ j=1,..,4 $, 
\be N_1^{(j)}(d,s) = \int_{0}^{b_1} db\ |b|^\frac{2-d}{2} 
    \ |R_4(b)|^\frac{d-3}{2} \times 
 \left( 1\ ; \ \frac{1}{b_2-b}\ ; \ \frac{1}{b_3-b}\ ; 
             \ \frac{1}{b_4-b} \right) \ , 
\labbel{eq:N1j} \ee 
with $ N_1^{(1)}(d,s) = N_1(d,s) $, 
and correspondingly the following 4 {\it integration by parts identities} 
\begin{align} 
    \int_{0}^{b_1} db \left( \delta(b-b_1)-\delta(b) 
                                 - \frac{d}{db}\right) 
    & |b|^\frac{2-d}{2}\ |R_4(b)|^\frac{d-1}{2}            \nonumber\\ 
    & \times \left[ (b_1-b) 
    \left( 1\ ; \ \frac{1}{b_2-b}\ ; \ \frac{1}{b_3-b}\ ; 
             \ \frac{1}{b_4-b} \right) \right] = 0 \ . 
\labbel{eq:b2bp} \end{align} 
By following the same procedure as for the previous case, one finds that, 
thanks to the identities Eq.s(\ref{eq:b2bp}), all the $b_3$ and $b_4$ 
derivatives of the {\it auxiliary integrals} Eq.(\ref{eq:N1j}), 
hence also the $s$ derivative of $ M_1(d,s) $, can be expressed in terms 
of those 4 {\it auxiliary integrals} only -- and then one can also check that 
$ M_1(d,s) $ does satisfy the 4th order equation Eq.(\ref{eq:ahe}). 
\par 
Similarly, the case of 
\be  N_5(d,s) = \int_{b_4}^{b_5=\infty} db 
                \ |b|^\frac{2-d}{2}\ |R_4(b)|^\frac{d-3}{2} \ , 
\labbel{eq:N5} \ee 
can be treated by introducing the 4 {\it auxiliary integrals}, 
for $ j=1,..,4$, 
\be N_5^{(j)}(d,s) = \int_{b_4}^{\infty} db\ |b|^\frac{2-d}{2} 
    \ |R_4(b)|^\frac{d-3}{2} \times 
 \left( 1\ ; \ \frac{1}{b-b_1}\ ; \frac{1}{b-b_2}\ ; 
             \ \frac{1}{b-b_3}\ ; \right) \ , 
\labbel{eq:N5j} \ee 
with  $ N^{(1)}_5(d,s) = N_5(d,s) [B$, 
and correspondingly the following 4 {\it integration by parts identities} 
\be \int_{b_4}^{\infty} db\ \frac{d}{db}\left[ 
    \ |b|^\frac{2-d}{2}\ |R_4(b)|^\frac{d-1}{2} \times\ (b-b_4) 
 \left( 1\ ; \ \frac{1}{b-b_1}\ ; \ \frac{1}{b-b_2}\ ; 
             \ \frac{1}{b-b_3} \right) \right] = 0 \ , 
\labbel{eq:b5bp} \ee 
where the vanishing end-point contributions were dropped for ease of 
typing. \par 
By following the above described procedure, the $s$-derivatives of 
$ M_5(d,s) $, can be expressed in terms of the 4 {\it auxiliary integrals} 
of Eq.(\ref{eq:N5j}) -- and then one cal also check that
$ M_5(d,s) $ does satisfy the 4th order equation Eq.(\ref{eq:ahe}). 
\par 
The last case to consider is 
\be N_0(d,s) = \int_{b_{-1}=-\infty}^{0} db
                \ |b|^\frac{2-d}{2}\ |R_4(b)|^\frac{d-3}{2} 
             = \int_0^\infty dc 
                \ c^\frac{2-d}{2}\ |R_4(-c)|^\frac{d-3}{2} \ . 
\labbel{eq:N0} \ee 
We proceed as  in the previous cases, with the {\it auxiliary integrals}, 
for $ j=1,..,4$, 
\be N_0^{(j)}(d,s) = \int_0^{\infty} dc\ c^\frac{2-d}{2}
    \ |R_4(-c)|^\frac{d-3}{2} \times
 \left( 1\ ; \ \frac{1}{c+b_1}\ ; \frac{1}{c+b_2}\ ;
             \ \frac{1}{c+b_3}\ ; \right) \ ,
\labbel{eq:N0j} \ee 
with $ N_0^{(1)}(d,s) = N_0(d,s) $, and the 
{\it integration by parts identities} 
\be \int_0^{\infty} dc\ \frac{d}{dc}\left[ c^\frac{2-d}{2} 
    \ |R_4(-c)|^\frac{d-3}{2} \times 
 \left( 1\ ; \ c\ ; \ \frac{1}{c+b_3}\ ; \ \frac{1}{c+b_4} 
             \right) \right] = 0 \ . 
\labbel{eq:b0bp} \ee 
Thanks to the identities, we are able to express the $s$-derivatives of 
Eq.(\ref{eq:N0}) and of $ M_0(d,s) $, Eq.(\ref{eq:Mi}) in terms of the 
{\it master integrals} of Eq.(\ref{eq:N0j}) -- and to check finally 
that also $ M_0(d,s) $ does satisfy the 4th order equation 
Eq.(\ref{eq:ahe}). 
\par 
Summarising, we have shown that all the 6 functions $ M_i(d,s), i=0,1,..,5 $ 
defined in Eq.s(\ref{eq:MNi},\ref{eq:Ni}) are solutions of the homogeneous 
equation Eq.(\ref{eq:ahe}), 
\be D_4(d,s) M_i(d,s) = 0 \ , \hskip1cm i=0,1,..,5 \ . 
\labbel{eq:D4Mi} \ee 
But the differential operator 
$ D_4(d,s) $ appearing in the equation is a 4th order operator, 
hence the differential equation has only 4 linearly independent 
solutions. We were however not able to show, so far, that for arbitrary $ d $, 
4 of those 6 solutions are indeed independent. In the following sections we 
will consider the cases of $ d=2,3,4 $, and we will show that in those cases 
4 of the solutions are indeed linearly independent, as expected. 
We further, again for $ d= 2,3,4 ,$ we show that the knowlegde of the 
solutions allows to factorise the 4th order operator $ D_4(d,s) $. 
\section{Derivatives and differential equation of the $b$ integrals 
for $d=2$} \labbel{sec:DD2} 
\setcounter{equation}{0} 
\numberwithin{equation}{section} 
We study in this section the solutions of Eq.(\ref{eq:D4Mi}) at $ d = 2 $. 
As the six solutions $ M_i(d,s) $, discussed in the above 
Section, are well defined for $ d $ in the range $ 1 < d < 8/3 $, 
and $ d=2 $ is in that range, we can write at once 
\be D_4(2,s)\ M_i(2,s) = 0 \ ,  \hskip8mm  i = 0,1,..,5 \ . 
\labbel{eq:ahe2} \ee 
where the functions $ M_i(2,s) $ correspond to the $ d=2 $ value of 
$ M_i(d,s) $, see Eq.s(\ref{eq:MNi},\ref{eq:Ni}). 
\par 
As a first comment, at $ d=2 $ the factor $ s^{\frac{2-d}{2}} $ in 
front of the $ b $ integral in Eq.(\ref{eq:MNi}) and the factor 
$ |b|^{\frac{2-d}{2}} $ within the integrand in Eq.(\ref{eq:Ni}) 
are both equal to 1, with an obvious simplification of the resulting 
expressions, which become 
\begin{align} M_i(2,s) &= N_i(2,s) \ , \labbel{eq:MNi2} \\ 
  N_i(2,s) &= \int_{b_{i-1}}^{b_i} 
                        db\ \frac{1}{\sqrt{|R_4(b)|}} 
  = \int_{b_{i-1}}^{b_i} db 
  \ \frac{1}{\sqrt{|(b-b_1)(b-b_2)(b-b_3)(b-b_4)|}} \ . \labbel{eq:Mi2} 
\end{align} 
\par 
The $ N_i(2,s) $ are elliptic integrals; their expression in terms of the 
complete and incomplete elliptic intregrals by means of the Legendre 
changes of variable are discussed in Appendix \ref{app:d2}. In this 
Section, in agreement with the rest of the paper, we discuss their 
properties by using directly the $b$-integral representation. 
For better exploiting the absence of the $ |b|^{\frac{2-d}{2}} $ and of 
the related absence of $ b = 0 $ as a sensible point, we will not use 
the $ d=2 $ limit of the results obtained in the previous Section, 
but will look for simpler identities and formulas. 
\par 
To start with, we want to evaluate the $s$-derivatives of the six 
$b$-integrals, beginning from the integrals whose integration limits 
correspond to two vanishing points of $ R_4(b) $, Eq.(\ref{eq:R4}). 
As a first case, we consider 
\be N_3(2,s) = \int_{b_2}^{b_3} db\ \frac{1}{\sqrt{|R_4(b)|}} 
             = \int_{b_2}^{b_3} db 
             \ \frac{1}{\sqrt{(b-b_1)(b-b_2)(b_3-b)(b_4-b)}} \ . 
\labbel{eq:M32} \ee 
For reasons which will be apparent later, we introduce the 
{\it partner integral} 
\be N^{(1)}_3(2,s) = \int_{b_2}^{b_3} db\ \frac{1}{\sqrt{|R_4(b)|}} 
                   \times \frac{1}{b-b_1} \ , \labbel{eq:M321} \ee 
and the two {\it integration by parts identities} 
\begin{align} 
 & \int_{b_2}^{b_3} db\ \frac{d}{db}\left[ 
 \frac{\sqrt{(b-b_2)(b_3-b)}}{\sqrt{(b-b_1)(b_4-b)}} \right] = 0 \ , 
                                                   \nonumber\\ 
 & \int_{b_2}^{b_3} db\ \frac{d}{db}\left[ 
 \frac{\sqrt{(b-b_2)(b_3-b)}}{\sqrt{(b-b_1)(b_4-b)}} 
          \times\ \frac{1}{b-b_1} \right] = 0 \ , \labbel{eq:M32bp} 
\end{align} 
where the vanishing end-point contributions were dropped. 
\par 
Working out the above expressions is elementary; without writing 
explicitly the results for simplicity, let us just say that they 
establish two identities between the four integrals 
\be \int_{b_2}^{b_3} db\ \frac{1}{\sqrt{|R_4(b)|}} \times \left[ 
    1\ ;\ \frac{1}{b-b_1}\ ;\ \frac{1}{b_4-b}\ ; 
        \ \frac{1}{(b-b_1)^2} \right] \ , \labbel{eq:M32id} \ee 
and the two identities can be used, when needed, to express the last two 
integrals in terms of the first two.
\par 
As already observed in the previous Section, the $s$-derivatives can be 
obtained from the $b_3$ and $b_4$ derivatives, see Eq.s(\ref{eq:bi}) 
for the definitions of the $b_i$. For $ N_3(2,s) $ and 
$ N^{(1)}_3(2,s) $ taking the $b_4$ derivative is straighforward 
(it gives just a factor $ 1/(b_4-b) $, as 
$b_4$ lies outside of the integration range), but that is not the 
case for the $b_3$ derivative; indeed, given the presence of $b_3$ as an 
end point, the na\"ive differentiation of the integrand is not allowed here 
(at variance with the previous section, in which $ d> 3$ was assumed; 
see however the comments following Eq.(\ref{eq:bmb3})). \par 
To overcome the probem let us introduce the function 
$ \atan{\sqrt{(b-b_2)/(b_3-b)}} $, whose derivatives in $b$ and $b_3$ are 
\begin{align} 
   \frac{d}{db} \left( \atan{\sqrt{\frac{b-b_2}{b_3-b}}} \right) &= 
        +\frac{1}{2}\ \frac{1}{\sqrt{(b-b_2)(b_3-b)}}\ , \labbel{eq:atdb}\\ 
 \frac{d}{db_3} \left( \atan{\sqrt{\frac{b-b_2}{b_3-b}}} \right) &= 
        -\frac{1}{2} \ \frac{b-b_2}{b_3-b_2} 
                \ \frac{1}{\sqrt{(b-b_2)(b_3-b)}}\ . 
\labbel{eq:atdb3} 
\end{align} 
By using Eq.(\ref{eq:atdb}) we can integrate by parts the factor 
$1/\sqrt{(b-b_2)(b_4-b)}$ in Eq.s(\ref{eq:M32},\ref{eq:M321}), obtaining 
\begin{align} 
   N_3(2,s) =&\ \frac{\pi}{\sqrt{(b_4-b_3)(b_3-b_1)}} \nonumber\\ 
            +& \int_{b_2}^{b_3} db 
            \ \left(\atan{\sqrt{\frac{b-b_2}{b_3-b}}} \right) 
            \ \frac{1}{\sqrt{(b-b_1)(b_4-b)}} 
            \left( \frac{1}{b-b_1} - \frac{1}{b_4-b} \right) \ , 
\labbel{eq:M32a} \\ 
  N^{(1)}_3(2,s) =&\ \frac{1}{b_3-b_1} 
                     \frac{\pi}{\sqrt{(b_4-b_3)(b_3-b_1)}} \nonumber\\ 
            +& \int_{b_2}^{b_3} db 
            \ \left(\atan{\sqrt{\frac{b-b_2}{b_3-b}}} \right) 
            \ \frac{1}{\sqrt{(b-b_1)(b_4-b)}} \times \left[ 
     \frac{3}{(b-b_1)^2} - \frac{1}{b_4-b_1} 
            \left( \frac{1}{b-b_1} + \frac{1}{b_4-b} \right) \right] \ . 
\labbel{eq:M321a} 
\end{align} 
The na\"ive differentiation in $ b_3 $ is now allowed, and gives 
\begin{align} 
  \frac{d}{db_3} N_3(2,s) =& \ \frac{1}{b_3-b_2} 
     \int_{b_2}^{b_3} db\ \frac{1}{\sqrt{|R_4(b)|}} 
     \times \biggl[ -1 + \frac{1}{2}(b_2-b_1)\times\frac{1}{b-b_1} 
                      + \frac{1}{2}(b_4-b_2)\times\frac{1}{b_4-b} \biggr] \ , 
\nonumber\\ 
  \frac{d}{db_3} N^{(1)}_3(2,s) =& \frac{1}{b_3-b_2} 
      \int_{b_2}^{b_3} db\ \frac{1}{\sqrt{|R_4(b)|}} 
  \times \biggl[ - \biggl( \frac{3}{2}\ + \frac{1}{2}\ \frac{b_2-b_1}{b_4-b_1}
                   \biggr)\times \frac{1}{b-b_1}  \nonumber\\ 
   & \hspace{32mm} + \frac{1}{2}\ \frac{b_4-b_2}{b_4-b_1}\times\frac{1}{b_4-b} 
      \ + \frac{3}{2}(b_2-b_1)\times \frac{1}{(b-b_1)^2}          \biggr] \ . 
   \labbel{eq:M321b} 
\end{align} 
The two expressions on the {\it r.h.s.} of Eq.(\ref{eq:M321b}) depend 
on the 4 integrals listed in Eq.(\ref{eq:M32id}); as already remarked, 
by using the Eq.s(\ref{eq:M32bp}), the integrals 
with the factors $ 1/(b_4-b), 1/(b-b_1)^2 $ can be expressed in terms 
of only the two integrals $ N_3(2,s) $ and $ N^{(1)}_3(2,s) $, defined by 
Eq.s(\ref{eq:M32},\ref{eq:M321}). \par 
As already observed, the allowed na\"ive differentiation with respect to $b_4$ 
of $ N_3(2,s) $ and $ N^{(1)}_3(2,s) $ generates only an extra factor 
$ 1/(b_4-b) $, so that, again, also the $b_4$ derivatives of 
$ N_3(2,s) $ and $ N^{(1)}_3(2,s) $ can be expressed in terms of 
the two same functions $ N_3(2,s) $ and $ N^{(1)}_3(2,s) $ only. 
\par 
Recalling Eq.(\ref{eq:dsNi}) and collecting results, one can then express 
the first order $s$-derivatives of $ N_3(2,s) $ and $ N^{(1)}_3(2,s) $ in 
terms of $ N_3(2,s) $ and $ N_3^{(1)}(2,s) $ only, obtaining two 
equations of the form 
\begin{align} 
        \frac{d}{ds} N_3(2,s) &= A(s)\ N_3(2,s) + B(s)\ N_3^{(1)}(2,s) \ , 
                                                    \nonumber\\ 
  \frac{d}{ds} N_3^{(1)}(2,s) &= C(s)\ N_3(2,s) + D(s)\ N_3^{(1)}(2,s) \ , 
                                                    \labbel{eq:M321sd} \ , 
\end{align} 
where the quantities $ A(s), B(s), C(s), D(s) $ are rational expressions 
in $s$ which we don't write for space reasons. 
By iterating the procedure, one can express also all the higher order 
$s$-derivatives of $ N_3(2,s) $ in terms of $ N_3(2,s) $ and 
$ N^{(1)}_3(2,s) $ only; finally, one can also verify by brute (algebraic) 
force that $ M_3(2,s) = N_3(2,s) $, see Eq.(\ref{eq:MNi2}), and its 
first 4 $s$-derivatives do satisfy the 
4th order equation Eq.(\ref{eq:ahe2}), as expected. 
\par 
But we can also rewrite the first of Eq.s(\ref{eq:M321sd}) as 
\be N_3^{(1)}(2,s) = - \frac{A(s)}{B(s)}\ N_3(2,s) 
                     + \frac{1}{B(s)}\ \frac{d}{ds} N_3(2,s) \ , 
\labbel{eq:M321sda} \ee 
expressing $ N_3^{(1)}(2,s) $ in terms of $ N_3(2,s) $ and 
$ dN_3(2,s)/ds $. 
\par 
We have already seen that all the $s$-derivatives of $ N_3(2,s) $ can 
be expressed in terms of $ N_3(2,s) $ and $ N_3^{(1)}(2,s) $ only; 
by using Eq.(\ref{eq:M321sda}) for replacing $ N_3^{(1)}(2,s) $ with 
$ dN_3(2,s)/ds $, we have that the 2nd and higher $s$-derivatives of 
$ N_3(2,s) $ can be expressed in terms of $ N_3(2,s) $ and 
$ dN_3(2,s)/ds $ only. In particular, the expression of the 2nd order 
derivative $ d^2N_3(2,s)/ds^2 $ in terms of $ N_3(2,s) $ and 
its 1st derivative $ dN_3(2,s)/ds $ is nothing but a 2nd order 
differential equation in $s$ for $ N_3(2,s) = M_3(2,s) $, which can be 
written as 
\be D_2(2,s)\ N_3(2,s) = 0 \ , \labbel{eq:D2M3} \ee 
where $ D_2(2,s) $ is a 2nd order differential operator, whose 
explicit expression is given in Eq.(\ref{eq:D22}). 
\par 
A similar procedure can be followed for 
\be N_2(2,s) = \int_{b_1}^{b_2} db\ \frac{1}{\sqrt{|R_4(b)|}} 
             = \int_{b_1}^{b_2} db 
             \ \frac{1}{\sqrt{(b-b_1)(b_2-b)(b_3-b)(b_4-b)}} \ , 
\labbel{eq:M22} \ee 
using as {\it partner integral} 
\be N^{(1)}_2(2,s) = \int_{b_1}^{b_2} db\ \frac{1}{\sqrt{|R_4(b)|}} 
     \times \frac{1}{b_4-b}\ , \labbel{eq:M221} \ee 
and as {\it integration by parts identities} the relations 
\be \int_{b_1}^{b_2} db\ \frac{d}{db}\left[ 
 \frac{\sqrt{(b-b_1)(b_2-b)}}{\sqrt{(b_3-b)(b_4-b)}} \times 
 \left( 1\ ; \frac{1}{b_4-b} \right) \right] = 0 \ . \nonumber 
\ee 
The direct, na\"ive differentiation ({\ie} without a preliminary integration 
by parts) of the integrals Eq.(\ref{eq:M22},\ref{eq:M221}) with respect 
to $b_3$ and $b_4$ is allowed, and, as in the case of $ M_3(2,s) $, one finds 
that also $ M_2(2,s) = N_2(2,s) $ satisfies, besides the 4th order equation 
(\ref{eq:ahe2}), also the 2nd order equation 
\be D_2(2,s)\ M_2(2,s) = 0 \ , \labbel{eq:D2M2} \ee 
where $ D_2(2,s,m_i^2) $ is the same as in Eq.(\ref{eq:D2M3}). 
\par 
The above approach can also be extended to 
\be N_4(2,s) = \int_{b_3}^{b_4} db\ \frac{1}{\sqrt{|R_4(b)|}} 
             = \int_{b_3}^{b_4} db 
             \ \frac{1}{\sqrt{(b-b_1)(b_2-b)(b-b_3)(b_4-b)}} \ , 
\labbel{eq:M42} \ee 
using as {\it partner integral} 
\be N^{(1)}_4(2,s) = \int_{b_3}^{b_4} db\ \frac{1}{\sqrt{|R_4(b)|}} 
     \times \frac{1}{b-b_1}\ , \labbel{eq:M421} \ee 
and as {\it integration by parts identities} the relations 
\be \int_{b_3}^{b_4} db\ \frac{d}{db}\left[ 
 \frac{\sqrt{(b-b_3)(b_4-b)}}{\sqrt{(b-b_1)(b-b_2)}} \times 
 \left( 1\ ; \frac{1}{b-b_1} \right) \right] = 0 \ . \labbel{eq:M421a} 
\ee 
Similarly to the case of $ N_2(2,s) $, for carrying out the $b_3$ and 
$b_4$ derivatives a preliminary integration by parts, 
involving this time the function 
$ \atan{\sqrt{(b-b_3)/(b_4-b)}} $, is needed; 
as in the case of $ M_3(2,s) $ and  $ M_2(2,s) $, one 
finds that $ M_4(2,s) = N_4(2,s) $ satisfies both the 4th order equation 
(\ref{eq:ahe2}) and the 2nd order equation 
\be D_2(2,s)\ M_4(2,s) = 0 \ , \labbel{eq:D2M4} \ee where 
$ D_2(2,s) $ is again the same differential operator as in 
Eq.s(\ref{eq:D2M3}) and (\ref{eq:D2M2}). 
\par 
To our knowledge, Eq.s(\ref{eq:D2M3},\ref{eq:D2M2},\ref{eq:D2M4}) 
were first obtained in the paper~\cite{Adams:2014vja}. 
\par 
As $ D_2(2,s) $ is a 2nd order differential 
operator, it admits only two linearly independent solutions, so a 
relation between the three solutions $ N_2(2,s) = M_2(2,s) $, 
$ N_3(2,s) = M_3(2,s) $ and $ N_4(2,s) = M_4(2,s) $ must exist. 
To find it, consider the contour integral 
\be \oint_{\mathfrak C} db \frac{1}{\sqrt{R_4(b)}} \nonumber \ee 
in the complex $b$ plane, where $ R_4(b) $ is the usual 4th order polynomial 
of Eq.(\ref{eq:R4}) (note the presence of $ \sqrt{R_4(b)} $ and not 
of $ \sqrt{|R_4(b)|} $ as in all the other so far considered integrals) 
and the contour $\mathfrak C$ is a (big) circle around the origin, of radius 
$ C $ with $ C \gg (b_i, i=1,..,4) $;  
as for large $b$ the function $ 1/\sqrt{R_4(b)} $ behaves as $ 1/b^2 $, 
the above contour integral vanishes 
\be \oint_{\mathcal C} db \frac{1}{\sqrt{R_4(b)}} = 0 \ . \labbel{eq:dbC} \ee 
Let us specify the analytic function $ {1}/{\sqrt{R_4(b)}} $ by 
giving it the real (and positive) value $ {1}/{\sqrt{|R_4(b)}|} $ in 
the real interval $ b_2 < b < b_3 $, two cuts along the two real intervals 
$ b_1 < b < b_2 $ and $ b_3 < b < b_4 $, and the real negative value 
$ \left( - {1}/{\sqrt{|R_4(b)}|} \right) $ along the remaining parts of the 
real $b$ axis, $ - \infty < b < b_1 $ and $ b_4 < b < +\infty $. 
By shrinking the big circle into two closed contours around the cuts, one 
arrives (up to an overall constant) at the equation  
\be N_2(2,s) - N_4(2,s) = 0 \ , \labbel{eq:M2M4} \ee 
which shows that two of the solutions are not linearly independent. 
(Strictly speaking, one should still show that two of them, say $ N_2(2,s) $ 
and $ N_3(2,s) $, are indeed independent, see below). 
\par 
We then consider the integrals whose integration limits do not correspond to 
the zeroes of $ R_4(b) $, starting from 
\be N_5(2,s) = \int_{b_4}^{\infty}db\ \frac{1}{\sqrt{|R_4(b)|}} 
             = \int_{b_4}^{\infty} db 
             \ \frac{1}{\sqrt{(b-b_1)(b-b_2)(b-b_3)(b-b_4)}} \ , 
\labbel{eq:M52} \ee 
(recall $ b_5 = \infty $), 
using as {\it partner integral} 
\be N^{(1)}_5(2,s) = \int_{b_4}^{\infty} db\ \frac{1}{\sqrt{|R_4(b)|}} 
     \times \frac{1}{b-b_1}\ , \labbel{eq:M521} \ee 
and as {\it integration by parts identities} the relations 
\begin{align} \int_{b_4}^{\infty} db\ \frac{d}{db}\left[ 
 \frac{\sqrt{(b-b_3)(b-b_4)}}{\sqrt{(b-b_1)(b-b_2)}} \times 
 \left( 1\ ;\ \frac{1}{b-b_1} \right) \right] &= \lim_{b\to\infty} 
 \left[
 \frac{\sqrt{(b-b_3)(b-b_4)}}{\sqrt{(b-b_1)(b-b_2)}} \times
 \left( 1\ ;\ \frac{1}{b-b_1} \right) \right] \nonumber\\ 
 &= \left[\ 1\ ;\ 0 \right] \ . \labbel{eq:M521a} 
\end{align}  
\par 
Note that, at variance with the previous cases, see for instance 
Eq.(\ref{eq:M421a}), the right hand side of the identities (\ref{eq:M521a}) 
is not zero, as the integration limits do not correspond to the zeroes of 
the integrand. We can however follow closely the above described 
procedures, expressing in that way all the $s$-derivatives of 
$ N_5(2,s) $ in terms of the two $b$-integrals $ N_5(2,s) $ and 
$ N^{(1)}_5(2,s) $ plus a bunch of additional algebraic terms, originated 
by the non vanishing right hand side of Eq.(\ref{eq:M521a}). 
\par 
The procedure followed for $ N_5(2,s) $ can be used also for the two 
$b$ integrals 
\begin{align} 
  N_0(2,s) &= \int_{-\infty}^{0}db\ \frac{1}{\sqrt{|R_4(b)|}} 
   = \int_{-\infty}^{0}db\ \frac{1}{\sqrt{(b_1-b)(b_2-b)(b_3-b)(b_4-b)}} 
                                                     \ , \nonumber\\ 
  N_1(2,s) &= \int_{0}^{b_1}db\ \frac{1}{\sqrt{|R_4(b)|}} 
   = \int_{0}^{b_1}db\ \frac{1}{\sqrt{(b_1-b)(b_2-b)(b_3-b)(b_4-b)}} 
                                                     \ , \labbel{eq:M012} 
\end{align} 
introducing as {\it partner integrals} 
\begin{align} 
  N^{(1)}_0(2,s) &= \int_{-\infty}^{0}db\ \frac{1}{\sqrt{|R_4(b)|}} 
                  \times \frac{1}{b_4-b}             \ , \nonumber\\ 
  N^{(1)}_1(2,s) &= \int_{0}^{b_1}db\ \frac{1}{\sqrt{|R_4(b)|}} 
                  \times \frac{1}{b_4-b}             \ . \labbel{eq:M012a} 
\end{align} 
Skipping details, one finds that, as in the case of $ N_5(2,s) $, all 
the $s$-derivatives of $ N_0(2,s) $ and $ N_1(2,s) $ can be expressed in 
terms of the pairs of functions $ N_0(2,s), N^{(1)}_0(2,s) $ and 
$ N_1(2,s), N^{(1)}_1(2,s) $, plus a bunch of additional algebraic terms. 
\par 
By evaluating all the $s$-derivatives in that way, one finds as expected 
that the three 
functions $ M_0(2,s) = N_0(2,s)\ $, 
          $ M_1(2,s) = N_1(2,s)\ $ and 
          $ M_5(2,s) = N_5(2,s)\ $ satisfy the 4th order 
equation (\ref{eq:ahe2}), but do not satisfy the 2nd order equation of 
Eq.(\ref{eq:D2M3}), satisfied by $ M_2(2,s), M_3(2,s), 4_4(2,s) $. 
\par 
Indeed, an explicit calculation gives 
\begin{align} 
    D_2(2,s) N_0(2,s) &= R_0(s,m_i^2) \ , \nonumber\\ 
    D_2(2,s) N_1(2,s) &= R_1(s,m_i^2) \ , \nonumber\\ 
    D_2(2,s) N_5(2,s) &= R_5(s,m_i^2) \ , \labbel{eq:D2M015} 
\end{align} 
where $ R_0(s,m_i^2), R_1(s,m_i^2), R_5(s,m_i^2) $, generated by the 
additional algebraic terms present in the expressions of the $s$-derivatives, 
are polynomials in $s$ and in the masses, and are found to satisfy the 
relation 
\be R_0(s,m_i^2) + R_1(s,m_i^2) + R_5(s,m_i^2) = 0 
                                    \ . \labbel{eq:R015} \ee 
For definiteness, the explicit expressions of $ R_0(s,m_i^2) $ and 
$ R_1(s,m_i^2) $ are 
\begin{align} 
 R_0(s,m_i^2) &= ( 2m_2^2 + 2m_3^2 - 4m_1^2 )\ s^3        \nonumber\\ 
 &+ ( 12m_1^4 - 6m_2^4 - 6m_3^4 
      - 14m_1^2 m_2^2 - 14m_1^2m_3^2 + 28m_2^2m_3^2 )\ s^2  \nonumber\\ 
 &+ ( 6 m_2^6 + 6 m_3^6 - 12 m_1^6 - 16 m_1^2 m_2^4 - 16 m_1^2 m_3^4 
     + 22 m_1^4 m_2^2 + 22 m_1^4 m_3^2                      \nonumber\\ 
 & \hspace{13mm}    - 6 m_2^2 m_3^4 - 6 m_2^4 m_3^2 )\ s    \nonumber\\ 
 & - 2 m_1^2 m_2^2 m_3^4 - 2 m_1^2 m_2^4 m_3^2 + 2 m_1^2 m_2^6 
   + 2 m_1^2 m_3^6 + 4 m_1^4 m_2^2 m_3^2 + 6 m_1^4 m_2^4  
   + 6 m_1^4 m_3^4                                          \nonumber\\ 
 & - 10 m_1^6m_2^2 - 10 m_1^6 m_3^2 + 4 m_1^8 - 2 m_2^8 - 2 m_3^8 
   + 8 m_2^2 m_3^6 - 12 m_2^4 m_3^4 + 8 m_2^6 m_3^2 \ ,     \labbel{eq:R0} 
\end{align} 
\begin{align} 
 R_1(s,m_i^2) &=  3( m_1^2 - m_2^2 )\ s^3 
    + 3( 3 m_2^4 - 3 m_1^4 + 7 m_1^2 m_3^2 - 7 m_2^2 m_3^2 )\ s^2 \nonumber\\ 
 &  + ( 9 m_1^6 - 9 m_2^6 + 5 m_1^2 m_3^4 - 5 m_2^2 m_3^4
       + 14 m_2^4 m_3^2 - 14 m_1^4 m_3^2 
       + 19 m_1^2 m_2^4 - 19 m_2^2 m_1^4 )\ s                     \nonumber\\ 
  & + 3( m_2^8 - m_1^8 + m_1^2 m_3^6 - m_2^2 m_3^6 
        + 2 m_1^6 m_2^2 - 2m_2^6 m_1^2 + 3 m_2^4 m_3^4 - 3 m_1^4 m_3^4
                                                                  \nonumber\\ 
  & \hspace{4mm} + 3 m_1^6 m_3^2 - 3 m_2^6 m_3^2 + m_1^2 m_2^4 m_3^2 
        - m_1^4 m_2^2 m_3^2 ) \ .                           \labbel{eq:R1} 
\end{align} 
\par 
The polynomial $ R_1(s,m_i^2) $ satisfies (obviously) a 1st order 
differential equation, which can be written as 
\be D_{1a}(2,s) R_1(s,m_i^2) = 0 \ ,           \labbel{eq:D21aR1} \ee 
where $ D_{1a}(2,s) $ is the 1st order differential operator 
\be D_{1a}(2,s) = R_1(s,m_i^2)\frac{d}{ds} 
                      - \left( \frac{dR_1(s,m_i^2)}{ds} \right) \ , 
\labbel{eq:D2D1a} \ee 
\par 
The polynomial $ R_0(s,m_i^2) $, which is different from $ R_1(s,m_i^2) $, 
cannot satisfy the same homogeneous 1st order equation (\ref{eq:D21aR1}) 
as $ R_1(s,m_i^2) $; indeed, an explicit calculation gives 
\be D_{1a}(2,s) R_0(s,m_i^2) = R_{0a}(s,m_i^2) \ , 
\labbel{eq:D21aR0} \ee 
where $ R_{0a}(s,m_i^2) $, whose expression is given by 
\be R_{0a}(s,m_i^2) = R_1(s,m_i^2)\frac{dR_0(s,m_i^2)}{ds} 
                    - \frac{dR_1(s,m_i^2)}{ds} R_0(s,m_i^2) \ , 
\labbel{eq:R0a} \ee 
is again a polynomial in $s$ and the masses (which we don't list here 
to save space). 
\par 
The polynomial $ R_{0a}(s,m_i^2) $ satisfies (obviously) another 
1st order differential equation, which can be written as 
\be D_{1b}(2,s) R_{0a}(s,m_i^2) = 0 \ ,       \labbel{eq:D21bR0a} \ee
where $ D_{1b}(2,s) $ is the 1st order differential operator 
\be D_{1b}(2,s) = R_{0a}(s,m_i^2)\frac{d}{ds}
                - \left( \frac{dR_{0a}(s,m_i^2)}{ds} \right) \ ,
\labbel{eq:D2D1b} \ee 
\par 
As it is easily verified, 
\be D_4(2,s) = \mathcal{K} 
               \ D_{1b}(2,s)\ D_{1a}(2,s)\ D_2(2,s)\ , 
\labbel{eq:D42} \ee 
where $ D_4(2,s) $ is the same operator as in Eq.(\ref{eq:ahe2}), 
\begin{equation} 
\mathcal{K} = -{8s^2}\times 
  \frac{ 5s^2 + 2(m_1^2 + m_2^2 + m_3^2) \ s - 7R_2(m_1^2,m_2^2,m_3^2) } 
       { 3s^2 - 2(m_1^2 + m_2^2 + m_3^2) \ s -  R_2(m_1^2,m_2^2,m_3^2) } 
  \times 
  \frac{1}{ R_{0a}(s,m_i^2) }\ \frac{1}{ R_{1}(s,m_i^2) } \labbel{eq:K} 
\end{equation} 
and 
\be R_2(m_1^2,m_2^2,m_3^2) = m_1^4+m_2^4+m_3^4 
    - 2m_1^2m_2^2 - 2m_2^2m_3^2 - 2m_3^2m_1^2 \ . \labbel{eq:R2m123} \ee 
\par 
Eq.(\ref{eq:K}) completes the factorisation of $ D_4(2,s) $ in the 
product of lower order differential operators with rational coefficients. 
\par 
It is to be noted that $ D_4(2,s) $ and $ D_2(2,s) $, see 
Eq.s(\ref{eq:D4a}) and Eq.(\ref{eq:D22}) for their explicit expessions, 
are symmetrical under the exchange of the three masses, but 
that is not the case for $ D_{1a}(2,s) $ and $ D_{1b}(2,s) $. 
\par 
Let us further observe that the factorisation of a differential operator 
of a given order into the product of differential operators of lower order 
is in general not unique, as shown for instance by the following elementary 
example 
$$ \left( \frac{d}{dx} + \frac{1}{x} \right) 
   \left( \frac{d}{dx} - \frac{1}{x} \right) = \frac{d^2}{dx^2} \ . $$ 
Indeed, an equivalent but different factorization of $ D_4(2,s) $ 
can be obtained by exchanging the order of $ R_0(s,m_i^2) $, 
$ R_1(s,m_i^2) $ and $ R_5(s,m_i^2) $ in the previous derivation. 
\par 
We will now look for a relationship of linear dependence 
between the three integrals $ N_0(2,s), N_1(2,s) $ and $ N_5(2,s) $, which 
do not satisfy the equation with the 2nd order operator $ D_2(2,s) $, 
and the three integrals $ N_2(2,s)$, $ N_3(2,s) $ and $ N_4(2,s) $, 
which do satisfy that equation. 
In analogy with Eq.(\ref{eq:dbC}) we can write 
\be \oint_{\mathfrak C} db \frac{1}{\sqrt{-R_4(b)}} = 0 \ , 
    \labbel{eq:dbCa} \ee 
where $ {\mathfrak C} $ is again the (big) circle around the origin, but 
this time we specify the analytic function $ 1/\sqrt{-R_4(b)} $ by assigning 
it real and positive value in the real interval $ b_1 < b < b_2 $, 
real (negative) value in the interval $ b_3 < b < b_4 $ and three 
cuts along the three real intervals $ -\infty < b < b_1 $, 
$ b_2 < b < b_3 $ and $ b_4 < b < +\infty $. By shrinking the big circle 
into three closed contours around the cuts, and observing that 
the interval $ -\infty < b < b_1 $ contributes to $ N_0(2,s) $ 
plus $ N_1(2,s) $, the interval $ b_2 < b < b_3 $ to 
$ - N_3(2,s) $ and $ b_4 < b < b_5 = +\infty $ to $ N_5(2,s) $, 
one arrives (up to an overall factor) at the equation 
\be N_0(2,s) + N_1(2,s) - N_3(2,s) + N_5(2,s) 
    = 0 \ , \labbel{eq:M0135} \ee 
which gives the expected linear relation between the various 
$b$-integrals. 
\par 
As a last argument for the discussion of the actual linear independence of 
4 of the 6 $b$-integrals, we evaluate them in the limit of large $ s=E^2 $, 
keeping a few leading terms in the $ 1/E $ expansion. 
Since $ b_3=(E-m_3)^2 $ and $ b_4 = (b+m_3)^2 $, Eq.(\ref{eq:bi}), 
the expansion in $ 1/E $ is equivalent to the expansion in 
$ 1/b_3, 1/b_4 $ with $b_3, b_4 $ larger than $ b_1, b_2 $. 
\par 
In $ N_1(2,s) $, see Eq.(\ref{eq:M012}), the integration range is 
$ 0 < b < b_1 $, 
the integration variable $ b $ is always smaller than $ b_3,b_4 $, so 
that we can expand $ \sqrt{b_3-b} $ in $ b/b_3 $,\ $ \sqrt{b_4-b} $ 
in $ b/b_4 $ 
and then integrate analytically the resulting terms which contain only 
the two square roots $ \sqrt{b_1-b}, \sqrt{b_2-b} $; the same argument 
applies to $ N_2(2,s) $, where $ b_1 < b < b_2 $. 
\par 
In $ N_4(2,s) $ the integration range is $ b_3 < b < b_4 $, so that 
$ b $ is always larger than $ b_1, b_2 $; we can therefore expand 
$ \sqrt{b-b_1} $ in $ b_1/b $, $ \sqrt{b-b_2} $ in $ b_2/b $ and then 
integrate analytically the resulting terms, which contain only the two 
square roots $ \sqrt{b-b_3}, \sqrt{b_4-b} $; the same argument applies to 
$ N_5(2,s) $, where $ b_4 < b < +\infty $. 
\par 
The case of $ N_3(2,s) $ is slightly more complicated, as the 
integration range, $ b_2 < b < b_3 $, includes both small and large values 
of the variable $ b $. For that reason, as a first step we split the range 
in two parts, the {\it small} $b$ range $ b_2 < b < B $ and the {\it large} 
$b$ range $ B < b < b_3 $, with $ b_2 \lless B \lless b_3 $, and we treat 
separately the two ranges. \par 
In the {\it small} $b$ range we expand $ \sqrt{b_3-b} $ and $ \sqrt{b_4-b} $ 
in $ b/b_3 $ and $ b/b_4 $, we then integrate analytically from 
$ b=b_2 $ up to $ b=B $ the terms so obtained, which contain only the two 
square roots $ \sqrt{b-b_1}, \sqrt{b-b_2} $, and finally we expand the 
analytical result in $ E $ and $ B $ (up to $1/B$, for the sake of later 
checks). 
In the {\it large} $b$ range we expand $ \sqrt{b-b_1}, \sqrt{b-b_2} $ in 
$ b_1/b,\ b_2/b $ we then integrate analytically from $ b=B $ up to $ b=b_3$ 
the terms so obtained, which contain only the two square roots 
$ \sqrt{b_3-b}, \sqrt{b_4-b} $, and finally we expand the analytical result 
in $ E $ and $ B $ \par 
As a concluding step we sum the contributions from the two $ b $ ranges, 
checking the cancellation of the dependence on $ B $ (as the result 
corresponds to the term of zeroth order in $ B $, for a stronger check we 
verify the disappearance of the $ B $-dependence up to the order $ 1/B $ 
included, although the terms in $ 1/B $ are anyhow irrelevant in the limit 
of large $ B $). 
\par 
Finally, the expansion of the remaining integral $ N_0(2,s) $ can 
be obtained by following the same procedure used for $ N_3(2,s) $. 
\par 
Let us indicate by $ N_k^{(a)}(2,s), k=0,..,5 $, the first 5 terms of 
the asymptotic expansions in $ 1/E^2 $, for large $ s = E^2 $, of the 6 
$ b $-integral $ N_k(2,s) $, starting from the leading term 
$ 1/E^2 $ up to the term $ 1/E^{10} $ included. The results of the 
above described calculations can be written as        
\begin{align} 
 N_0^{(a)}(2,s) &= B_0(E,m_i)\ln\left(\frac{E^2}{m_1^2}\right) 
                                             + B_1(E,m_i) \ , \nonumber\\ 
 N_1^{(a)}(2,s) &= B_0(E,m_i)\ln\left(\frac{m_1}{m_2}\right) 
                                             + B_2(E,m_i) \ , \nonumber\\ 
 N_2^{(a)}(2,s) &= B_0(E,m_i)\pi                      \ , \nonumber\\ 
 N_3^{(a)}(2,s) &= B_0(E,m_i)\ln\left(\frac{E^3}{m_1 m_2 m_3}\right) 
                                             + B_3(E,m_i) \ , \nonumber\\ 
 N_4^{(a)}(2,s) &= B_0(E,m_i)\pi                      \ , \nonumber\\ 
 N_5^{(a)}(2,s) &= B_0(E,m_i)\ln\left(\frac{E}{m_3}\right) 
                                             + B_4(E,m_i) \ , \labbel{eq:Nia} 
\end{align} 
with 
\begin{align} 
 B_0(E,m_i) &= 1/E^2 + \bigl( m_1^2 + m_2^2 + m_3^2 \bigr)/E^4   \nonumber\\ 
  &+ \bigl( 4 m_1^2 m_2^2 + 4 m_1^2 m_3^2 + 4 m_2^2 m_3^2 
                            + m_1^4 + m_2^4 + m_3^4 \bigr)/E^6   \nonumber\\ 
  &+ \bigl( 36 m_1^2 m_2^2 m_3^2 + 9 m_1^2 m_2^4 + 9 m_1^2 m_3^4 
     + 9 m_1^4 m_2^2 + 9 m_1^4 m_3^2 + 9 m_2^2 m_3^4 
     + 9 m_2^4 m_3^2                                             \nonumber\\ 
  & \hspace{5mm} + m_1^6   + m_2^6 + m_3^6 \bigr)/E^8            \nonumber\\ 
  &+ \bigl( 144 m_1^2 m_2^2 m_3^4 
          + 144 m_1^2 m_2^4 m_3^2 
          + 144 m_1^4 m_2^2 m_3^2 
           + 36 m_1^4 m_2^4   
           + 36 m_2^4 m_3^4 
           + 36 m_1^4 m_3^4 
\nonumber\\ 
  & \hspace{5mm} 
           + 16 m_1^2 m_2^6 
           + 16 m_1^2 m_3^6 
           + 16 m_1^6 m_2^2 
           + 16 m_1^6m_3^2 
           + 16 m_2^2 m_3^6 
           + 16 m_2^6 m_3^2          \nonumber\\ 
  & \hspace{5mm} + m_1^8 + m_2^8 + m_3^8 \bigr)/E^{10} \ ,  \labbel{eq:B0Emi} 
\end{align} 
\begin{align}
 B_1(E,m_i) &= \bigl( - 2 m_2^2 - 2 m_3^2 \bigr)/E^4 
    + \bigl( - 4 m_1^2 m_2^2 - 4 m_1^2 m_3^2 - 12 m_2^2 m_3^2 
             - 3 m_2^4 - 3 m_3^4 \bigr)/E^6                      \nonumber\\ 
  &+ \bigl( - 60 m_1^2 m_2^2 m_3^2 - 15 m_1^2 m_2^4 - 15 m_1^2 m_3^4 
     - 6 m_1^4 m_2^2 - 6 m_1^4 m_3^2 - 33 m_2^2 m_3^4            \nonumber\\ 
  & \hspace{5mm} - 33 m_2^4 m_3^2 - 11 m_2^6/3 
                                  - 11 m_3^6/3 \bigr)/E^8        \nonumber\\ 
  &+ \bigl( - 312 m_1^2 m_2^2 m_3^4 - 312 m_1^2 m_2^4 m_3^2 
            - 104 m_1^2 m_2^6/3 - 104 m_1^2 m_3^6/3 - 168 m_1^4 m_2^2 m_3^2 
                                                                 \nonumber\\ 
  & \hspace{5mm} - 42 m_1^4 m_2^4 - 42 m_1^4 m_3^4 - 8 m_1^6 m_2^2 
   - 8 m_1^6 m_3^2 - 200 m_2^2 m_3^6/3 - 200 m_2^6 m_3^2/3 
\nonumber\\ 
  & \hspace{5mm} - 150 m_2^4 m_3^4         
                 - 25 m_2^8/6 - 25 m_3^8/6 \bigr)/E^{10}\ , \labbel{eq:B1Emi} 
\end{align} 
\begin{align}
 B_2(E,m_i) &= \bigl( - m_1^2 + m_2^2 \bigr)/E^4 
    + \bigl( - 4 m_1^2 m_3^2 + 4 m_2^2 m_3^2 
             - 3 m_1^4/2 + 3 m_2^4/2 \bigr)/E^6                  \nonumber\\ 
  &+ \bigl( 9 m_1^2 m_2^4/2 - 9 m_1^2 m_3^4 - 9 m_1^4 m_2^2/2 
     - 27 m_1^4 m_3^2/2 - 11 m_1^6/6 + 9 m_2^2 m_3^4             \nonumber\\ 
  & \hspace{5mm} + 27 m_2^4 m_3^2/2 + 11 m_2^6/6 \bigr)/E^8      \nonumber\\ 
  &+ \bigl( + 72 m_1^2 m_2^4 m_3^2 + 40 m_1^2 m_2^6/3 
     - 16 m_1^2 m_3^6  - 72 m_1^4 m_2^2 m_3^2 - 54 m_1^4 m_3^4 
     - 40 m_1^6 m_2^2/3                                          \nonumber\\ 
  & \hspace{5mm}  - 88 m_1^6 m_3^2/3 
     - 25 m_1^8/12 + 16 m_2^2 m_3^6 + 54 m_2^4 m_3^4 
     + 88 m_2^6 m_3^2/3 + 25 m_2^8/12    \bigr)/E^{10}\ , \labbel{eq:B2Emi} 
\end{align} 
\begin{align}
 B_3(E,m_i) &= \bigl( - 2 m_1^2 - 2 m_2^2 - 2 m_3^2 \bigr)/E^4   \nonumber\\ 
  &+ \bigl( - 10 m_1^2 m_2^2 - 10 m_1^2 m_3^2 m_1^4 - 10 m_2^2 m_3^2 
            - 3 m_1^4 - 3 m_2^4 - 3 m_3^4 \bigr)/E^6             \nonumber\\ 
  &+ \bigl( - 90 m_1^2 m_2^2 m_3^2 - 27 m_1^2 m_2^4 - 27 m_1^2 m_3^4 
     - 27 m_1^4 m_2^2 - 27 m_1^4 m_3^2 - 27 m_2^2 m_3^4 
     - 27 m_2^4 m_3^2                                            \nonumber\\ 
  & \hspace{5mm} - 11 m_1^6/3 - 11 m_2^6/3 - 11 m_3^6/3 
                                             \bigr)/E^8          \nonumber\\ 
  &+ \bigl(      - 396 m_1^2 m_2^2 m_3^4 
                 - 396 m_1^2 m_2^4 m_3^2 
                 - 396 m_1^4 m_2^2 m_3^2 
                 - 164/3 m_1^2 m_2^6 
                 - 164/3 m_1^2 m_3^6                             \nonumber\\ 
  & \hspace{5mm} - 164/3 m_1^6 m_2^2 
                 - 164/3 m_1^6 m_3^2 
                 - 164/3 m_2^2 m_3^6 
                 - 164/3 m_2^6 m_3^2 
                 - 117 m_1^4 m_2^4 
                 - 117 m_1^4 m_3^4                               \nonumber\\ 
  & \hspace{5mm} - 117 m_2^4 m_3^4 
                 - 25 m_1^8/6 - 25 m_2^8/6 - 25 m_3^8/6 
\bigr)/E^{10} \ ,                                           \labbel{eq:B3Emi} 
\end{align} 
\begin{align}
 B_4(E,m_i) &= \bigl( - m_1^2 - m_2^2 \bigr)/E^4               \nonumber\\ 
  &+ \bigl(      - 10 m_1^2 m_2^2 - 10 m_1^2 m_3^2 - 10 m_2^2 m_3^2 
                 - 3 m_1^4 - 3 m_2^4 - 3 m_3^4 \bigr)/E^6        \nonumber\\ 
  &+ \bigl(      - 30 m_1^2 m_2^2 m_3^2 
                 - 33 m_1^2 m_2^4/2 
                 - 33 m_1^4 m_2^2/2 
                 -  3 m_1^2 m_3^4 
                 -  3 m_2^2 m_3^4                                \nonumber\\ 
  & \hspace{5mm} - 15 m_1^4 m_3^2/2 
                 - 15 m_2^4 m_3^2/2 
                 - 11 m_1^6/6 
                 - 11 m_2^6/6 \bigr)/E^8                         \nonumber\\ 
  &+ \bigl(      -  84 m_1^2 m_2^2 m_3^4 
                 - 156 m_1^2 m_2^4 m_3^2 
                 - 156 m_1^4 m_2^2 m_3^2 
                 - 100 m_1^2 m_2^6/3 
                 - 100 m_1^6 m_2^2/3                             \nonumber\\ 
  & \hspace{5mm} -   4 m_1^2 m_3^6 
                 -   4 m_2^2 m_3^6 
                 -  75 m_1^4 m_2^4 
                 -  21 m_1^4 m_3^4 
                 -  21 m_2^4 m_3^4 
                 -  52 m_1^6 m_3^2/3 
                 -  52 m_2^6 m_3^2/3                             \nonumber\\ 
  & \hspace{5mm} -  25 m_1^8/12 
                 -  25 m_2^8/12 \bigr)/E^{10} \ ,           \labbel{eq:B4Emi} 
\end{align} 

As the functions $ N_k(2,s) $ are solutions of the 4th differential equation 
(\ref{eq:ahe2}), the expansions $ N_k^{(a)}(2,s) $ can be seen as 
(actually redundant) boundary conditions at large $ s=E^2 $ which identify 
completely the corresponding solutions $ N_k(2,s) $. 
\par 
It is immediate to verify that they satisfy the relations 
\begin{align} 
 N_4^{(a)}(2,s) &= N_2^{(a)}(2,s) \ , \nonumber\\ 
 N_5^{(a)}(2,s) &= - N_0^{(a)}(2,s) - N_1^{(a)}(2,s) 
                       + N_3^{(a)}(2,s) \ ; \labbel{eq:N5a} 
\end{align} 
as any homogeneous relations valid for the boundary conditions 
$ N_k^{(a)}(2,s) $ holds also for the corresponding solutions $ N_k(2,s) $. 
Eq.s(\ref{eq:N5a}) imply the already obtained relations 
Eq.(\ref{eq:M2M4}) end Eq.(\ref{eq:M0135}). 
\par 
From the explicit epressions of Eq.s(\ref{eq:B1Emi}--\ref{eq:B4Emi}) 
one can also check that the 4 expansions $ N_k^{(a)}(2,s), k=0,..,3 $ 
are indeed linearly independent from each other, so that 
the corresponding 4 functions $ N_k(2,s), i=0,..,3 $ are also 
linearly independent of each other. 
\par 
Let us now briefly discuss the symmetry of the solutions under 
permutations of the three masses. To that aim, for evidentiating 
explicitly the dependence on the masses, we introduce a new family of 4 
functions $ N_k(2,s,m_1,m_2,m_3), k=0,..,3 $ defined by 
\be N_k(2,s,m_1,m_2,m_3) = N_k(2,s); \hspace{4mm} k=0,..,3 
\labbel{eq:Nim} \ee 
in other words, the $ N_k(2,s) $, with the already considered asymptotic 
expansions $ N_i^{(a)}(2,s) $, correspond to the permutation 
$ (m_1,m_2,m_3) $ of the masses, while $ N_k(2,s,m_2,m_3,m_1) $, for 
instance, correspond to the function whose asymptotic expansion is obtained 
by the permutation $ (m_2,m_3,m_1) $ of the masses in Eq.s(\ref{eq:Nia}) 
and (\ref{eq:B1Emi}--\ref{eq:B4Emi}). 
\par 
As already observed, the 4th order differential operator $ D_4(d,s) $ 
appearing in Eq.(\ref{eq:D4M}) is fully symmetrical in the 3 masses, and 
the same is obviously true for $ D_4(2,s) $, its value at $ d=2 $; 
therefore, if $ N_k(2,s,m_1,m_2,m_3), k=0,..,3 $ is a solution of the equation 
\be D_4(2,s)\ N_k(2,s,m_1,m_2,m_3) = 0 \ , \labbel{eq:D42Ni} \ee 
each of the 6 functions $ N_k(2,s,m_{p_1},m_{p_2},m_{p_3}) $, where 
$ (m_{p_1},m_{p_2},m_{p_3} ) $ is any of the 6 permutations of the masses, 
is also a solution of that equation. We obtain in that way a total of 
$ 4 \times 6 = 24 $ solutions; but there are only 4 linearly 
indendent solutions of the homogeneous equation fo the 4th order 
differential operator $ D_4(2,s) $, so that all the new functions 
introduced with the permutations of the masses must be linear combinations 
of the 4 solutions $ N_k(2,s), i=0,..,3 $. 
\par 
That is easily verified by the explicit form of the asymptotic expansions 
Eq.s(\ref{eq:Nia}) and (\ref{eq:B1Emi}--\ref{eq:B4Emi}). 
One finds immediately that 
$ N_2^{(a)}(2,s,m_1,m_2,m_3) $ and $ N_3^{(a)}(2,s,m_1,m_2,m_3) $ are 
fully symmetrical in the masses, so that they remain identical to 
$ N_2^{(a)}(2,s) $ and $ N_3^{(a)}(2,s) $ under the mass permutations. 
One further finds 
\begin{align} 
  N_0^{(a)}(2,s,m_1,m_3,m_2) &=    N_0^{(a)}(2,s) \ , \nonumber\\ 
  N_0^{(a)}(2,s,m_3,m_1,m_2) &= -2 N_0^{(a)}(2,s) 
                                -2 N_1^{(a)}(2,s) 
                                +2 N_3^{(a)}(2,s) \ , \nonumber\\ 
  N_0^{(a)}(2,s,m_2,m_3,m_1) &=    N_0^{(a)}(2,s) 
                                +2 N_1^{(a)}(2,s) \ , \labbel{eq:N0aa} 
\end{align} 
and 
\begin{align} 
  N_1^{(a)}(2,s,m_2,m_1,m_3) &=       N_1^{(a)}(2,s) \ , \nonumber\\ 
  N_1^{(a)}(2,s,m_3,m_1,m_2) &= \frac{3}{2} N_0^{(a)}(2,s) 
                                + N_1^{(a)}(2,s) 
                                - N_3^{(a)}(2,s)     \ , \nonumber\\ 
  N_1^{(a)}(2,s,m_3,m_2,m_1) &= \frac{3}{2} N_0^{(a)}(2,s) 
                                + 2 N_1^{(a)}(2,s) 
                                -   N_3^{(a)}(2,s)   \ . \labbel{eq:N1aa} 
\end{align} 
showing, as expected, that all the new expansions introduced with the 
permutations of the masses are indeed linear combinations of the 4 
functions $ N_k^{(a)}(2,s), k=0,..,3 $, and the same holds for all the 
functions $ N_k(2,s,m_{p_1},m_{p_2},m_{p_3}) $, which are therefore 
linear combinations of the 4 solutions $ N_k(2,s), i=0,..,3 $. 
\par 
A last argument to discuss is the equal mass limit $ m_1=m_2=m_3= m $. 
An elementary calulation shows that the equal mass limit of the 4th order 
differential operator $ D_4(2,s) $, of Eq.(\ref{eq:ahe2}), 
let us call it $ D_4(2,s,m) $ (without writing it explicitly for the 
sake of brevity), is still a 4th order 
operator, which therefore admits 4 independent homogeneous solutions. 
\par 
By calling $ N_i^{(a)}(2,s,m) $ and $ B_k(E,m) $ the the equal mass limit 
of $ N_i^{(a)}(2,s) $ and $ B_k(E,m_i) $ given by Eq.(\ref{eq:Nia}), 
one finds 
\begin{align} 
  N_0^{(a)}(2,s,m) &= \frac{2}{3} N_3^{(a)}(2,s,m) \ , \nonumber\\ 
  N_1^{(a)}(2,s,m) &= 0 \ ,                            \nonumber\\ 
  N_2^{(a)}(2,s,m) &= B_0(E,m)\pi \ ,                  \nonumber\\ 
  N_3^{(a)}(2,s,m) &= B_0(E,m)\ln\left(\frac{E^3}{m^3}\right) 
                    + B_3(E,m) \ ,                     \labbel{eq:Niam} 
\end{align} 
showing that only two of the 4 linearly independent asymptotic expansions 
of the solutions, say $ N_2^{(a)}(2,s,m) $ and $ N_3^{(a)}(2,s,m) $, 
are left, corresponding to two solutions only of the 4th order 
homogeneous equation of $ D_4(2,s,m) $, the equal mass limit of 
$ D_4(2,s) $. Two solutions are apparently missing. 
\par 
But by looking again at Eq.s(\ref{eq:Nia}) and 
(\ref{eq:B1Emi}--\ref{eq:B4Emi}), 
one finds that the large $ E $ expansions 
of the two {\it missing solutions} are provided by         
\begin{align} 
  \tilde{N}_0^{(a)}(2,s,m) &= \lim_{m_i\to m} \frac{m_1}{m_3-m_1} 
     \left[ \frac{2m_1}{m_2-m_1} N_1^{(a)}(2,s)
            + \frac{m_1}{m_3-m_1} 
              \left( 3N_0^{(a)}(2,s)-2N_3^{(a)}(2,s) \right) 
     \right] \  \nonumber\\ 
  &= - \frac{1}{E^2} -5\frac{m^2}{E^4} - 9\frac{m^4}{E^6} 
     - 13\frac{m^6}{E^8}- 17\frac{m^8}{E^{10}} \ , \nonumber\\ 
  \tilde{N}_1^{(a)}(2,s,m) &= \lim_{m_i\to m} \frac{m_1}{m_2-m_1} 
                                   N_1^{(a)}(2,s) \nonumber\\ 
  &= - \frac{1}{E^2} - \frac{m^2}{E^4} - \frac{m^4}{E^6} 
     - \frac{m^6}{E^8}- \frac{m^8}{E^{10}} \ , \labbel{eq:Nt01a} 
\end{align} 
where the equal mass limit implies the L'H\^opital's rule. The two 
{\it missing solutions} correspond then to two homogeneous solutions of 
Eq.(\ref{eq:ahe2}) at equal masses with the asymptotic expansions provided 
by Eq.s(\ref{eq:Nt01a}). 
\section{ The differential equation at $ d=3 $ and its solutions } 
\labbel{sec:DD3} 
\setcounter{equation}{0} 
\numberwithin{equation}{section} 
At $ d = 3 $, Eq.(\ref{eq:D4Mi}) becomes 
\be D_4(3,s) \ M_i(3,s) = 0 \ , \hskip5mm i = 0,1,..,5 \ .  
\labbel{eq:34th}  \ee 
where $ D_4(3,s) $, according to Eq(\ref{eq:D4}), is given by 
\begin{align}  
       D_4(3,s) &= C_4(3,s) \frac{d^4}{ds^4} 
                 + C_3(3,s) \frac{d^3}{ds^3} \nonumber\\ 
                &+ C_2(3,s) \frac{d^2}{ds^2} 
                 + C_1(3,s) \frac{d}{ds} 
                 + C_0(3,s)\ . \labbel{eq:D43} 
\end{align} 
In particular one finds 
\begin{align} 
    C_4(3,s) &= 32 s^3  \mathcal{D}(s) 
          \bigl[s^2 - R_2(m_1^2,m_2^2,m_3^2) \bigr] \ , \nonumber\\ 
    C_0(3,s) &= 0 \ ,  \labbel{eq:C403} 
\end{align} 
see Eq.(\ref{eq:DB3}) for $ \mathcal{D}(s) $ and Eq.(\ref{eq:R2m123}) 
for $ R_2(m_1^2,m_2^2,m_3^2) $. 
\par 
The na\"ive $ d = 3 $ limit of the functions $ M_i(3,s) $ appearing 
in Eq.(\ref{eq:34th}), according to Eq.s(\ref{eq:MNi},\ref{eq:Ni}), 
is particularly simple and reads 
\be  M_i(3,s) = \frac{1}{E} N_i(3,s) \ , \labbel{eq:Mi3E} \ee 
with $ s = E^2 $ as usual, and 
\be  N_i(3,s) = \int_{b_i-1}^{b_i} db\ \frac{1}{\sqrt{|b|}} \ . 
\labbel{eq:MNi3s} \ee 
But for $ i=0 $ and $ i = 5 $ the na\"ive $ d = 3 $ limit cannot 
be used, because it would give the non convergent integrals 
\begin{align} 
  N_0(3,s) &= \int_{-\infty}^{0} db\ \frac{1}{\sqrt{|b|}} \ , \nonumber\\ 
  N_5(3,s) &= \int_{b_4}^{+\infty} db\ \frac{1}{\sqrt{|b|}} \ . 
\labbel{eq:N053} 
\end{align} 
To overcome the inconvenience, 
one must go back to the definitions of $ N_0(d,s), N_5(d,s) $, which 
converge for $ d < 8/3 $, as already remarked in Section \ref{sec:Mc}, 
and continue them analytically to $ d > 3 $. 
\par 
According to Eq.(\ref{eq:Ni}), the definition of $ N_0(d,s) $ is 
\be N_0(d,s) = 
   \int_{-\infty}^{0} db\ |b|^{\frac{2-d}{2}} |R_4(b)|^{\frac{d-3}{2}} \ , 
\labbel{eq:N0x} \ee 
which can be rewritten as 
$$ N_0(d,s) = 
   \int_{-\infty}^{0} db\ |b|^{\frac{3}{2}d-5} 
      \left[\left|\frac{R_4(b)}{b^4}\right|^{\frac{d-3}{2}}\right]  $$ 
By integrating by parts the factor $ |b|^{\frac{3}{2}d-5} $ one obtains 
\be N_0(d,s) = -\frac{2}{3d-8}\int_{-\infty}^0 db\ |b|^{\frac{3}{2}d-4}
   \ \frac{d}{db}\left[\left|\frac{R_4(b)}{b^4}\right|^{\frac{d-3}{2}}
                    \right] \ , \labbel{eq:N03a}
\ee
an integral representation which converges for $ d < 10/3 $, hence valid 
at $ d = 3 $.  
\par 
By a similar argument, for $ i=5 $ one can arrive at the analytical 
continuation 
\be N_5(d,s) = -\frac{2}{3d-8}\int_{b_4}^{\infty} db
             \ (b-b_4)^{\frac{3}{2}d-4} \
   \ \frac{d}{db}\left[\left(\frac{b}{b-b_4}\right)^{-\frac{d-2}{2}}
                       \left(\frac{R_4(b)}{(b-b_4)^4}\right)^{\frac{d-3}{2}}
                 \right] \ , \labbel{eq:N53a}
\ee
which is again converging for $ d = 3 < 10/3 $. 
\par 
The na\"ive $ d = 3 $ limit of Eq.s(\ref{eq:N03a},\ref{eq:N53a}) is now 
allowed, and the $b$-integrations can be carried out for all the 
$ N_i(3,s) $ integrals. 
The results are, with $ s = E^2 $, 
\begin{align} 
 N_0(3,s) &= 0                \ , \nonumber\\ 
 N_1(3,s) &= 2(m_1-m_2)       \ , \nonumber\\ 
 N_2(3,s) &= 4 m_2            \ , \nonumber\\ 
 N_3(3,s) &= 2(E-m_1-m_2-m_3) \ , \nonumber\\ 
 N_4(3,s) &= 4 m_3            \ , \nonumber\\ 
 N_5(3,s) &= - 2(E+m_3)       \ , \labbel{eq:Ni3a} 
\end{align} 
It is apparent that only two of the above functions are linearly independent. 
Considering the corresponding solutions of Eq.(\ref{eq:34th}) 
$ M_i(3,s) = N_i(3,s)/E $, see Eq.(\ref{eq:Mi3E}), one finds that 
those solutions are combinations of just two linearly 
independent solutions, say, for definiteness 
\begin{align} F_1(3,s) &= 1 \ , \nonumber\\ 
              F_2(3,s) &= \frac{1}{E} \ .    \labbel{eq:Fk2} 
\end{align} 
Let us note here, in passing, that inserting $ F_2(3,s) $ and its 
derivatives in the equation 
\be D_4(3,s) F_2(3,s) = 0 \ , \labbel{eq:D4F2} \ee 
and by keeping $ D_4(3,s) $ in the form of Eq.(\ref{eq:D43}), 
one immediately obtains the relation 
\be                C_1(3,s) - \frac{3}{2s}    C_2(3,s) 
  + \frac{15}{4s^2}C_3(3,s) - \frac{105}{8s^3}C_4(3,s) = 0 \ . 
\labbel{eq:C12343} \ee 
\par 
As often observed, the 4th order equation Eq.(\ref{eq:34th}) is expected to 
have 4 linearly indepent solutions, while Eq.(\ref{eq:Fk2}) gives only two. 
To complete the list of the solutions, let us go back from Eq.(\ref{eq:34th}),
valid at $ d=3 $, to the equation for generic $ d $, which we rewrite here as 
\be D_4(d,s)\ G(d,s) = 0 \ , \labbel{eq:D4G} \ee 
and expand everything around $ d = 3 $ up to first order in $ (d-3) $ 
\begin{align} 
    D_4(d,s) &= D_4(3,s) + (d-3) D_4(3,1,s) + ... \ , \nonumber\\ 
      G(d,s) &=   G(3,s) + (d-3)   G(3,1,s) + ... \ . \labbel{eq:D4G3d} 
\end{align} 
Correspondingly, Eq.(\ref{eq:D4G}) splits into an equation at $ d=3 $ and an 
equation at 1st order in $ (d-3) $ 
\begin{align} 
      & D_4(3,s)\ G(3,s) = 0 \ ,                        \labbel{eq:D4G30} \\ 
      & D_4(3,s)\ G(3,1,s) + D_4(3,1,s)\ G(3,s) = 0 \ . \labbel{eq:D4G31} 
\end{align} 
If 
\be G(3,s) = 0 \ ,  \labbel{eq:G30} \ee 
Eq.(\ref{eq:D4G31}) becomes 
\be D_4(3,s)\ G(3,1,s) \ = 0 \ ,  \labbel{eq:G31} \ee 
{\it i.e.} if $ G(3,s) = 0 $, then $ G(3,1,s) $ is a solution of the 
equation with the differential operator $ D_4(3,s) $. 
\par 
We will then look for linear combinations $ G_k(d,s) $ of the functions 
$ M_i(d,s) $ whose $ d = 3 $ limits vanish, $ G_k(3,s) = 0 $, so that 
the first order terms in $ (d-3) $ of their expansions around $ d = 3 $, 
$ G_k(3,1,s) $, are solutions of Eq.(\ref{eq:G431}). 
\par 
To that aim, we expand also the functions $ M_i(d,s) $ and $ N_i(d,s) $ 
around $ d = 3 $ up to first order in $(d-3)$ 
\begin{align} 
     M_i(d,s) &= M_i(3,s) + (d-3) M_i(3,1,s) \ , \nonumber\\ 
     N_i(d,s) &= N_i(3,s) + (d-3) N_i(3,1,s) \ , \labbel{eq:MNi3} 
\end{align} 
which according to Eq.(\ref{eq:MNi}) gives 
\begin{align} 
 M_i(3,s) &= \frac{1}{E} N_i(3,s) \ , \labbel{eq:MNi3x}\\ 
 M_i(3,1,s) &= \frac{1}{E} \bigg( -\ln(E)N_i(3,s) + N_i(3,1,s) \bigg) \ . 
                                      \labbel{eq:MNi31} 
\end{align} 
It is then easily found, on account of Eq.s(\ref{eq:Ni3a}), that 
the following 4 linear combination $ G_k(d,s), k=1,..,4 $ 
of the 6 functions $ M_i(d,s) $ 
\begin{align} 
  G_1(d,s) &= M_1(d,s) + M_2(d,s) + M_3(d,s) + M_4(d,s) + M_5(d,s) 
                                                        \ , \nonumber\\ 
  G_2(d,s) &= 2 m_2 M_1(d,s) - (m_1-m_2)M_2(d,s)        \ , \nonumber\\ 
  G_3(d,s) &= m_3 M_2(d,s) - m_2 M_4(d,s)               \ , \nonumber\\ 
  G_4(d,s) &= 2m_3 M_3(d,s) + (m_1+m_2+2m_3)M_4(d,s) + 2m_3 M_5(d,s) \ , 
\labbel{eq:Gkd} \end{align} 
do satisfy, by construction, the condition 
\be G_k(3,s) = 0 \ . \hspace{2cm} k=1,..,4   \labbel{eq:Gk3} \ee 
Therefore, according to Eq.s(\ref{eq:G30},\ref{eq:G31}) 
the functions $ G_k(3,1,s) $ satisfy Eq.(\ref{eq:G31}), namely 
\be D_4(3,s)\ G_k(3,1,s) = 0 \ . \hspace{2cm} k=1,..,4 \ . 
\labbel{eq:G431a} \ee 
\par 
The evaluation of the functions $ G_k(3,1,s) $, according to 
Eq.(\ref{eq:Gkd}) and Eq.(\ref{eq:MNi31})  requires the evaluation of 
the functions $ N_i(3,1,s) $; one finds 
\begin{align} 
 N_0(3,1,s) &= 2\pi( E+m_1 )                       \ ,  \nonumber\\ 
 N_1(3,1,s) &= 2\ m_1 \big( \ln(4 m_1) - 3 \big) 
             - 2\ m_2 \big( \ln(4 m_2) - 3 \big)          \nonumber\\ 
    &+ (E+m_1-m_2+m_3) \ln(E+m_1-m_2+m_3) \nonumber\\ 
    &+ (E+m_1-m_2-m_3) \ln(E+m_1-m_2-m_3) \nonumber\\ 
    &- (E-m_1+m_2+m_3) \ln(E-m_1+m_2+m_3) \nonumber\\ 
    &- (E-m_1+m_2-m_3) \ln(E-m_1+m_2-m_3) \labbel{eq:N131} 
\end{align} 
\begin{align} 
 N_2(3,1,s) &= 4\ m_2  \big( \ln(4m_2) - 3 \big) \nonumber\\ 
    &+ (E+m_1+m_2+m_3) \ln(E+m_1+m_2+m_3) \nonumber\\ 
    &+ (E+m_1+m_2-m_3) \ln(E+m_1+m_2-m_3) \nonumber\\ 
    &- (E+m_1-m_2+m_3) \ln(E+m_1-m_2+m_3) \nonumber \\
    &- (E+m_1-m_2-m_3) \ln(E+m_1-m_2-m_3) \nonumber \\
    &+ (E-m_1+m_2+m_3) \ln(E-m_1+m_2+m_3) \nonumber \\
    &+ (E-m_1+m_2-m_3) \ln(E-m_1+m_2-m_3) \nonumber \\
    &- (E-m_1-m_2+m_3) \ln(E-m_1-m_2+m_3) \nonumber \\
    &- (E-m_1-m_2-m_3) \ln(E-m_1-m_2-m_3) \ , \labbel{eq:N231} 
\end{align} 
\begin{align} 
 N_3(3,1,s) &= 2E\ \big( \ln(4E) - 3 \big) 
             - 2m_1\ \big( \ln(4m_1) - 3 \big) \nonumber\\ 
            &- 2m_2\ \big( \ln(4m_2) - 3 \big) 
             - 2m_3\ \big( \ln(4m_3) - 3 \big) \nonumber\\ 
    - & (E+m_1+m_2+m_3) \ln(E+m_1+m_2+m_3) \nonumber\\ 
    + & (E+m_1-m_2-m_3) \ln(E+m_1-m_2-m_3) \nonumber\\ 
    + & (E-m_1+m_2-m_3) \ln(E-m_1+m_2-m_3) \nonumber\\ 
    + & (E-m_1-m_2+m_3) \ln(E-m_1-m_2+m_3) \nonumber\\ 
    + &2(E-m_1-m_2-m_3) \ln(E-m_1-m_2-m_3) \ , \labbel{eq:N331} 
\end{align} 
\begin{align} 
 N_4(3,1,s) =& 4\ m_3 \big( \ln(4m_3) - 3 \big) \nonumber\\ 
    + & (E+m_1+m_2+m_3) \ln(E+m_1+m_2+m_3) \nonumber\\ 
    - & (E+m_1+m_2-m_3) \ln(E+m_1+m_2-m_3) \nonumber\\ 
    + & (E+m_1-m_2+m_3) \ln(E+m_1-m_2+m_3) \nonumber\\ 
    - & (E+m_1-m_2-m_3) \ln(E+m_1-m_2-m_3) \nonumber\\ 
    + & (E-m_1+m_2+m_3) \ln(E-m_1+m_2+m_3) \nonumber\\ 
    - & (E-m_1+m_2-m_3) \ln(E-m_1+m_2-m_3) \nonumber\\ 
    + & (E-m_1-m_2+m_3) \ln(E-m_1-m_2+m_3) \nonumber\\ 
    - & (E-m_1-m_2-m_3) \ln(E-m_1-m_2-m_3) \ , \labbel{eq:N431} 
\end{align} 
\begin{align}
 N_5(3,1,s) &= -2E \ln(4E) - 2m_3\ln(m_3)  \nonumber\\
     &- (E+m_1+m_2+m_3) \ln(E+m_1+m_2+m_3) \nonumber\\
     &- (E+m_1-m_2+m_3) \ln(E+m_1-m_2+m_3) \nonumber\\
     &- (E-m_1+m_2+m_3) \ln(E-m_1+m_2+m_3) \nonumber\\
     &- (E-m_1-m_2+m_3) \ln(E-m_1-m_2+m_3) \ . \labbel{eq:N531}
\end{align}
\par 

The explicit values 
of the $ G_k(3,1,s) $ are then easily obtained from the various definitions 
and the values of the functions $ N_i(3,1,s) $ listed above. For writing 
them down it is convenient to introduce the new function 
\begin{align} 
 F(3,s,m_1,m_2,m_3) 
    &= (m_1-m_2)(E+m_1+m_2+m_3) \ln(E+m_1+m_2+m_3)          \nonumber\\ 
    &+ (m_1-m_2)(E+m_1+m_2-m_3) \ln(E+m_1+m_2-m_3)          \nonumber\\ 
    &- (m_1+m_2)(E+m_1-m_2+m_3) \ln(E+m_1-m_2+m_3)          \nonumber\\ 
    &- (m_1+m_2)(E+m_1-m_2-m_3) \ln(E+m_1-m_2-m_3)          \nonumber\\ 
    &+ (m_1+m_2)(E-m_1+m_2+m_3) \ln(E-m_1+m_2+m_3)          \nonumber\\ 
    &+ (m_1+m_2)(E-m_1+m_2-m_3) \ln(E-m_1+m_2-m_3)          \nonumber\\ 
    &- (m_1-m_2)(E-m_1-m_2+m_3) \ln(E-m_1-m_2+m_3)          \nonumber\\ 
    &- (m_1-m_2)(E-m_1-m_2-m_3) \ln(E-m_1-m_2-m_3)     \ ,  \labbel{eq:F3} 
\end{align} 
which satisfies under permutation of the masses the following identities 
\begin{align} 
    & F(3,s,m_1,m_2,m_3) + F(3,s,m_2,m_1,m_3) = 0 \ , \nonumber\\ 
    & F(3,s,m_2,m_3,m_1) + F(3,s,m_3,m_2,m_1) = 0 \ , \nonumber\\ 
    & F(3,s,m_3,m_1,m_2) + F(3,s,m_1,m_3,m_2) = 0 \ , \nonumber\\ 
    & m_3F(3,s,m_1,m_2,m_3) + m_1F(3,s,m_2,m_3,m_1) 
                            + m_2F(3,s,m_3,m_1,m_2) = 0 \ ; \labbel{eq:F3a} 
\end{align} 
according to the above 4 identities 2 of the 6 functions, say for 
definiteness $ F(3,s,m_1,m_2,m_3) $ and $ F(3,s,m_2,m_3,m_1) $, can be 
taken as linearly independent. 
\par 
As $ M_i(3,1,s) = N_i(3,1,s)/E, i=0,1,..,5 $, 
the explicit expressions of the four functions $ G_k(3,1,s), k=1,..,4\ , $ 
are given by 
\be G_1(3,1,s) = - \frac{6}{E} (E+m_3) \ , 
\labbel{eq:G131} \ee 
\be G_2(3,1,s) = \frac{1}{E}\biggl[ 4m_1m_2\ln\left(\frac{m_1}{m_2}\right) 
                 - F(3,s,m_1,m_2,m_3) \biggr] \ , 
\labbel{eq:G231} \ee 
\be G_3(3,1,s) = \frac{1}{E}\biggl[ 4m_2m_3\ln\left(\frac{m_2}{m_3}\right) 
                - F(3,s,m_2,m_3,m_1) \biggr] \ , 
\labbel{eq:G331} \ee 
\begin{align}
    G_4(3,1,s) &= \frac{1}{E}\biggl[ -12m_3(E+m_3) 
                   - 4m_1m_3\ln\left(\frac{m_1}{m_3}\right)
                   - 4m_2m_3\ln\left(\frac{m_2}{m_3}\right) \nonumber\\ 
               &+ F(3,s,m_2,m_3,m_1) - F(3,s,m_3,m_1,m_2) \biggr] \ . 
\labbel{eq:G431} \end{align} 
\par 
The four functions $ G_k(3,1,s), k=1,..,4 $. see Eq.(\ref{eq:G431a}), are 
all solutions of Eq.(\ref{eq:G31}); we have already seen, Eq.(\ref{eq:Fk2}, 
two of the four linearly indepent solutions of the equation, it is not 
difficult, using in particular Eq.s(\ref{eq:F3a}), that two other 
independent solutions are given by 
\begin{align} 
        F_3(3,s) &= \frac{1}{E}\ F(3,s,m_1,m_2,m_3) \ , \nonumber\\ 
        F_4(3,s) &= \frac{1}{E}\ F(3,s,m_2,m_3,m_1) \ . \labbel{eq:F3k} 
\end{align} 
Summarising, we have found that the 4 linearly independent functions 
$ F_k(3,s), k=1,..,4 $, defined by Eq.s(\ref{eq:Fk2},\ref{eq:F3k}), 
can be choosen as a set of 4 linearly independent solutions of the equation 
\be D_4(3,s)\ F_k(3,s) = 0 \ . \labbel{eq:D4F3} \ee 
\par 
Once the four solutions are knwon, we can factorise the 4th order 
diferential operator $ D_4(3,s) $. \par 
To that aim, by following the same procedure followed in the previous 
Section, let us begin by introducing the two first order differential 
operators 
\begin{align} 
 D_a(3,s) &= \frac{d}{ds} \ , \nonumber\\ 
 D_b(3,s) &= 2s\ \frac{d}{ds} + 3 \ , \labbel{eq:D3ab} 
\end{align} 
such that 
\begin{align} 
  &           D_a(3,s)\ F_1(3,s) = 0 \ , \nonumber\\ 
  & D_b(3,s)\ D_a(3,s)\ F_2(3,s) = 0 \ . \labbel{eq:D3ab0} 
\end{align} 
An elementary calculation gives 
\be 
  D_b(3,s)\ D_a(3,s)\ F_3(3,s) = R_{3c}(s,m_i^2) \ , 
\labbel{eq:R3c} \ee 
where $ R_{3c}(s,m_i^2) $ is given by the rational expression 
\be 
 R_{3c}(3,s,m_i^2) = 8 \frac{m_1m_2(m_2^2-m_1^2)}{\sqrt{s}\ \mathcal{D}(s)} 
          \bigl[s^2 - 2(m_1^2+m_2^2-3m_3^2)s + R_2(m_1^2,m_2^2,m_3^2) 
          \bigr] \ ,                          \labbel{eq:dfR3c} 
\ee 
with $ \mathcal{D}(s) $ and $ R_2(m_1^2,m_2^2,m_3^2) $ given by 
Eq.(\ref{eq:DB3}) and Eq.(\ref{eq:R2m123}). \\ 
The rational expression $ R_{3c}(3,s,m_1^2) $ satisfies, by construction, the 
first order differential equation 
\be D_c(3,s) R_{3c}(3,s,m_1^2) = 0 \ , \labbel{eq:DcR3c} 
\ee 
where $ D_{c}(3,s) $ is the first order differential operator 
\begin{equation} 
 D_c(3,s) = R_{3c}(3,s,m_i^2)\frac{d}{ds} - 
            \bigg(\frac{dR_{3c}(3,s,m_i^2)}{ds}\bigg) ,   \labbel{eq:D3c} 
\end{equation} 
and $ dR_{3c}(3,s,m_i^2)/ds $, like $ R_{3c}(3,s,m_i^2) $  is also a 
rational expression. 
\par 
Another explicit calculation gives 
\be D_c(3,s)\ D_b(3,s)\ D_a(3,s)\ F_4(3,s) = R_{3d}(s,m_i^2) \ , 
\labbel{eq:R3d} \ee 
with 
\be 
 R_{3d}(3,s,m_i^2) = 512\frac{m_1m_2^2m_3}{s\mathcal{D}^2(s)} 
   (m_3^2-m_2^2)(m_3^2-m_1^2)(m_2^2-m_1^2) 
   \bigl[s^2 - R_2(m_1^2,m_2^2,m_3^2) \bigr] \ .    \labbel{eq:dfR3d} 
\ee 
\par 
Again, $ R_{3d}(3,s,m_i^2) $ satisfies by construction the first order 
differential equation 
\be D_d(3,s) R_{3d}(3,s,m_1^2) = 0 \ , \labbel{eq:DdR3d} 
\ee 
where $ D_{d}(3,s) $ is the first order differential operator 
\begin{equation} 
 D_d(3,s) = R_{3d}(3,s,m_i^2)\frac{d}{ds} - 
            \bigg(\frac{dR_{3d}(3,s,m_i^2)}{ds}\bigg) ,   \labbel{eq:D3d} 
\end{equation} 
and $ dR_{3d}(3,s,m_i^2)/ds $, like $ R_{3d}(3,s,m_i^2) $, is also a 
rational expression. 
\par 
The full factorisation of $ D_4(3,s) $ is now completed and reads 
\be 
 D_4(3,s) = \frac{1}{\mathcal{C}} 
            \ D_d(3,s) D_c(3,s) D_b(3,s) D_a(3,s) \ , \labbel{D43fact} 
\ee 
with        
\be {\mathcal{C}} = 
    256\frac{m_1^2 m_2^3 m_3}{s^3 \sqrt{s} \mathcal{D}^4(s)} 
    (m_2^2-m_1^2)^2 (m_3^2-m_2^2) (m_3^2-m_1^2) 
    \bigl[ s^2 - 2(m_1^2+m_2^2-3m_3^2)s + R_2(m_1^2,m_2^2,m_3^2) \bigr] \ . 
\labbel{eq:defC} \ee 
\section{The differential equation at d = 4 and its solutions} 
\labbel{sec:DD4} 
\setcounter{equation}{0} 
\numberwithin{equation}{section} 
In this Section we study the d = 4 limit of the 4th order differential 
equation Eq(\ref{eq:D4Mi}), which becomes 
\be D_4(4,s) \ M_i(4,s) = 0 \ . \hskip5mm i = 0,1,..,5 \ . 
\labbel{eq:44th}  \ee 
As in the $ d=3 $ case of the previous Section, the values at $ d=4 $ of 
the five coefficients $ C_i(d,s,m_i^2), i=0,1,..,4 $ which appear in 
$ D_4(d,s) $ can be immediately obtained from Eq(\ref{eq:D4}), 
\begin{align}  
 D_4(4,s) &= C_4(4,s) \frac{d^4}{ds^4} 
           + C_3(4,s) \frac{d^3}{ds^3} \nonumber\\ 
          &+ C_2(4,s) \frac{d^2}{ds^2} 
           + C_1(4,s) \frac{d}{ds} 
           + C_0(4,s)\ . \labbel{eq:D44} 
\end{align} 
In particular one finds 
\begin{align} 
    C_4(4,s) &= 8 s^3  \mathcal{D}(s) 
          \bigl[3s^2 - 2(m_1^2 m_2^2 m_3^2)s - R_2(m_1^2,m_2^2,m_3^2) 
                                                       \bigr] \ , \nonumber\\ 
    C_0(4,s) &= 0 \ ,  \labbel{eq:C404} 
\end{align} 
see Eq,(\ref{eq:DB3}) for $ \mathcal{D}(s) $ and Eq.(\ref{eq:R2m123}) 
for $ R_2(m_1^2,m_2^2,m_3^2) $. 
\par 
The first step for discussing the solutions of Eq.(\ref{eq:44th}) is 
obviously to look at the $ d=4 $ limit of the 6 functions defined in 
Eq.s(\ref{eq:MNi},\ref{eq:Ni}), 
which we recall here for the convenience of the reader 
\be M_i(d,s) = s^{\frac{2-d}{2}}N_i(d,s) \ , \labbel{eq:MNi4} \ee 
with 
\be N_i(d,s) = \int_{b_{i-1}}^{b_i} db 
            \ |b|^\frac{2-d}{2}\ |R_4(b)|^\frac{d-3}{2} \ , 
\labbel{eq:Ni4} \ee 
where the $ b_i, i=-1,0,..,5 $ specified in Eq.(\ref{eq:bi}). 
The na\"ive $ d = 4 $ limit of Eq.(\ref{eq:Ni4}) then reads 
\be N_i(4,s) = \int_{b_{i-1}}^{b_i} \frac{db}{|b|} 
                                  \ \sqrt{|R_4(b|)} \ . 
\labbel{eq:N44} \ee 
\par 
As already observed in Section 2, for $ i=0 $ and $i=5$ the above definitions 
of the $ N_i(d,s) $ converge at large $ |b| $ only if $ d < 8/3 $, further 
for $ i = 0,1 $ they converge at $ b=0 $ only if $ d < 4 $; for reaching 
$ d=4 $ an analytic continuation to $ d > 4 $ is needed. 
\par 
The analytic continuations can be obtained by suitable integrations by parts. 
For $ i = 0 $, with the obvious change $ b=-c $ of the integration variable, 
the analytic continuation can be written as 
\begin{align} N_0(d,s) &= \frac{-16}{3(d-4)^2(3d-8)(3d-10)} \nonumber\\ 
  &\times \int_0^\infty c^{-\frac{1}{2}(d-4)} 
    \frac{d}{dc} \bigg\{ c^{1+2(d-4)} 
    \frac{d}{dc} \bigg[ c^2 
    \frac{d}{dc} \bigg( c^2 
    \frac{d}{dc} \bigg( \frac{R_4(-c)}{c^4}\bigg)^{\frac{1}{2}(d-3)} 
    \bigg)\bigg]\bigg\} 
\labbel{eq:N0d4} 
\end{align} 
with $ R_4(-c) = (c+b_1)(c+b_2)(c+b_3)(c+b_4) $, according to 
Eq.(\ref{eq:R4}). 
The above equation can be expanded in Laurent series in $(d-4)$ 
\be N_0(d,s) = \frac{1}{(d-4)^2} N_0(4,-2,s) 
            + \frac{1}{(d-4)} N_0(4,-1,s) + N_0(4,0,s) + .... \ ; 
\labbel{N040} \ee 
an explicit, relatively simple calculation then gives 
\begin{align} N_0(4,-2,s) &= 0 \ , \nonumber\\ 
  N_0(4,-1,s) 
  &= \frac{4}{3}( - 2 m_1^2 + m_2^2 + m_3^2 ) E^2 \nonumber\\ 
  &+ \frac{4}{3}( + m_1^2 m_2^2 + m_1^2 m_3^2 - 2 m_2^2 m_3^2 ) . 
\labbel{eq:N04-1} 
\end{align} 
By expanding also $ M_0(d,s) = $ around $ d=4 $, 
$$ M_0(d,s) = \frac{1}{(d-4)^2} M_0(4,-2,s) 
            + \frac{1}{(d-4)}   M_0(4,-1,s) + .... \ , $$ 
Eq.(\ref{eq:MNi4}) gives as leading term of the expansion of $ M_0(d,s) $ 
\be M_0(4,-1,s) = \frac{1}{E^2}  N_0(4,-1,s) \ , 
\labbel{eq:M04} \ee 
with $ s = E^2 $, 
and correspondingly Eq.(\ref{eq:44th}) for $ i=0 $ becomes, again at leading 
order in $ (d-4) $ 
\be \frac{1}{(d-4)}\times D_4(4,s) M_0(4,-1,s) = 0 \ , \labbel{eq:D4M0-1} \ee 
which implies, by dropping the overall $s$-independent facor 
$ 1/(d-4) $, 
\be D_4(4,s) M_0(4,-1,s) = 0 \ . \labbel{eq:D44M0-1a} \ee 
Hence, 
\begin{align} M_0(4,-1,s) &= \frac{1}{E^2}  N_0(4,-1,s) \nonumber\\ 
  &= \frac{4}{3}( - 2 m_1^2 + m_2^2 + m_3^2 ) \nonumber\\ 
  &+ \frac{4}{3 E^2}( + m_1^2 m_2^2 + m_1^2 m_3^2 - 2 m_2^2 m_3^2 )\ , 
  \labbel{eq:M40-1} 
\end{align} 
is a non vanishing solution of Eq.(\ref{eq:44th}). 
\par 
For the $ i=1 $ and $ i= 5 $ cases of Eq.(\ref{eq:Ni4}) with the same notation
for the analytic continuations we find 
\begin{align} 
 N_1(4,-1,s) &= 
  ( - 2m_1^2 + 2 m_2^2 )E^2 + 2m_1^2 m_3^2 - 2 m_2^2 m_3^2 \ , \nonumber\\ 
 N_5(4,-1,s) &= 
    \frac{16}{3} (- m_1^2 - m_2^2 + 2 m_3^2 )E^2  \nonumber\\ 
 &+ \frac{16}{3} ( 2 m_1^2 m_2^2 - m_1^2 m_3^2 - m_2^2 m_3^2 ) \ , 
\labbel{eq:N154-1} 
\end{align} 
corresponding to the solutions 
\begin{align} 
 M_1(4,-1,s) &= ( - 2m_1^2 + 2 m_2^2 ) + 
             \frac{1}{E^2} (2m_1^2 m_3^2 - 2 m_2^2 m_3^2 \ , \nonumber\\ 
 M_5(4,-1,s) &= 
    \frac{16}{3}    (- m_1^2 - m_2^2 + 2 m_3^2 ) \nonumber\\ 
 &+ \frac{16}{3E^2} ( 2 m_1^2 m_2^2 - m_1^2 m_3^2 - m_2^2 m_3^2 ) \ , 
\labbel{eq:M154-1} 
\end{align} 
It is easily seen that the three above solutions, $ M_i(4,-1,s) , i=0,1,5 $ 
are linear comninations of just two linearly independent solutions, 
say, for definiteness 
\begin{align} F_1(4,s) &= 1 \ , \nonumber\\ 
              F_2(4,s) &= \frac{1}{E^2} \ . \labbel{eq:F412}
\end{align} 
\par 
Following the derivation of Eq.(\ref{eq:C12343}), by inserting $ F_2(4,s) $ 
and its derivatives in the equation 
\be D_4(4,s)\ F_2(4,s) = 0 \ , \labbel{eq:D44F2} \ee  
and by keeping $ D_4(4,s) $ in the form of Eq.(\ref{eq:D44}), one immediately 
obtains the relation 
\be                C_1(4,s,m_i^2) -    \frac{2}{s}C_2(4,s,m_i^2) 
    + \frac{6}{s^2}C_3(4,s,m_i^2) - \frac{24}{s^3}C_4(4,s,m_i^2) = 0 
\labbel{eq:C12344} \ee 
between the coefficients of Eq.(\ref{eq:D44}). 
\par 
Let us look now for the other two linearly independent solutions of the 
4th order equation Eq.(\ref{eq:44th}). 
\par 
For the evaluation of the $ d=4 $ limit of the other three integrals 
$ N_i(d,s), i=2,3,4 $ no analytic continuation in $ d $ is needed, and the 
straightforward use of Eq.(\ref{eq:N44}) is allowed. 
\par 
In order to discuss their linear dependence, in strict analogy with the 
discussion leading to Eq.(\ref{eq:M2M4}), let us consider the analytic 
function 
\be  f(b) = \frac{1}{b} \sqrt{R_4(b)} \ , \label{eq:phi}        \ee 
with $ R_4(b) $ defined by Eq.(\ref{eq:R4}), 
which has a pole at $ b = 0 $, a cut for $ b_1 < b < b_2 $ and another cut for 
$ b_3 < b < b_4 $. 
\par 
By integrating that function in the complex b plane along the circle 
$ b = C {\rm e}^{i\phi}, 0 < \phi < 2\pi$ in the limit of large 
$ C \gg E^2, m_i^2 $, expanding $ f(b) $ in powers of $ b $ and dropping 
terms of order $1/C$ in the result, one finds 
\be I_{\mathcal C} = \oint_{\mathcal C} db\ f(b) 
                   = 2i\pi\bigl[ (m_1^2 + m_2^2 - 2 m_3^2) E^2 
                     - 2 m_1^2 m_2^2  + m_1^2 m_3^2 + m_2^2 m_3^2  
                         \bigr]  \ .\ \labbel{eq:ointI} \ee 
On the other hand, by squeezing the circle into a thin, elongated rectangle 
with vertices 
$$ -C-i\eps, {\hskip2mm} +C-i\eps, {\hskip2mm} +C+i\eps, 
                                   {\hskip2mm} -C+i\eps, $$ 
one finds, in the $ C \eps \to 0 $ limit 
\begin{align} J_{\mathcal C} 
  &= \oint_{\tikz[baseline=0.6ex]\draw (0,0) rectangle (3mm,3pt);} 
     db f(b) 
   = \int_{-C}^{+C} db\biggl[ f(b-i\eps) - f(b+i\eps) \biggr] \nonumber\\ 
  &= 4i\pi (m_3^2 - m_2^2) (E^2 - m_1^2) 
   + 2i\int_{b_1}^{b_2} db \frac{1}{b} \sqrt{|R_4(b)|} 
   - 2i\int_{b_3}^{b_4} db \frac{1}{b} \sqrt{|R_4(b)|} \ , \labbel{eq:ointJ} 
\end{align} 
where the first term comes from the pole at $ b= 0 $, the other two from 
the two cuts. The equality of $ I_{\mathcal C} $ and $ J_{\mathcal C} $ 
gives, in the notation of Eq.s(\ref{eq:N04-1},\ref{eq:N154-1}), 
(\ref{eq:Ni4}) and \ref{N040}, 
\be N_4(4,0,s) = \frac{3}{2}\pi N_0(4,-1,s) - 2\pi N_1(4,-1,s) 
               + N_2(4,0,s) \ , \labbel{N440s} \ee 
showing the that $ N_4(4,0,s) $ is a linear combination of the integrals 
in the {\it r.h.s.}. The above relation is to be compared with 
Eq.(\ref{eq:M2M4}), the analog and simpler equation relating $ N_4(2,s) $ 
and $ N_2(2,s) $ at $ d=2 $. 
\par 
In order to complete (and check) the discussion of the linear 
dependence or independence of the six 
integrals $ N_i(d,s) $ at $ d = 4 $, one can consider, besides the simple 
explicit expressions of $ N_0(4,-1,s) $ and $ N_1(4,-1,s) $, also 
the expansions 
of $ N_2(4,s), N_3(4,s) $ and $ N_4(4,s) $ at large $ E $, up to a 
sufficiently high oder (say up to $ 1/E^{10} $ included, redundantly), which 
can be evaluated by the same techniques leading to 
Eq.s(\ref{eq:Nia})-(\ref{eq:B4Emi}) at $ d = 2 $. Without reporting the
results for the sake of brevity, let us just say that they provide a 
check of Eq.(\ref{N440s}) and show that the four linearly independent 
solutions 
of Eq.(\ref{eq:44th}) can be taken to be the following four functions 
\begin{align} 
   F_1(4,s) &= 1 \ , \nonumber\\ 
   F_2(4,s) &= \frac{1}{E^2} \ , \nonumber\\ 
   F_3(4,s) &= \frac{1}{E^2} N_2(4,s) \ , \nonumber\\ 
   F_4(4,s) &= \frac{1}{E^2} N_3(4,s) \ . \labbel{eq:Fk4} 
\end{align} 
Correspondingly, we can factorise the 4th order differential operator 
$ D_4(4,s) $ in the form 
\be D_4(4,s) = D_c(4,s) D_b(4,s) D_a(4,s) \ , 
    \labbel{eq:D44fact} \ee 
where 
\begin{align} D_a(4,s) &= \frac{d}{ds} \ , \nonumber\\ 
              D_b(4,s) &= \frac{d}{ds} + \frac{2}{s} \nonumber 
\end{align} 
are fixed by imposing 
\begin{align} D_a(4,s) F_1(4,s) &= 0 \ , \nonumber\\ 
     D_b(4,s) D_a(4,s) F_2(4,s) &= 0 \ , \nonumber  
\end{align} 
while $ D_c(4,s) $ is a second order differential operator 
\be D_c(4,s) = A_2(4,s) \frac{d^2}{ds^2} 
             + A_1(4,s) \frac{d}{ds} + A_0(4,s) \ , \labbel{eq:Dc4} 
\ee 
with 
\begin{align} 
 A_2(4,s) &= C_4(4,s) \ , \nonumber\\ 
 A_1(4,s) &= C_3(4,s) - \frac{2}{s}C_4(4,s) \ , \nonumber\\ 
 A_0(4,s) &= C_2(4,s) - \frac{2}{s}C_3(4,s) 
                      + \frac{8}{s^2}C_4(4,s,m_i^2) \ , 
\labbel{eq:A2104} 
\end{align} 
as immediately found by comparing Eq.s(\ref{eq:D44fact}) and (\ref{eq:Dc4}) 
with Eq.(\ref{eq:D44}). \par 
Note that the above relations do not involve $ C_1(4,s) $, which 
indeed, according to Eq.(\ref{eq:C12344}), can be expressed in terms of 
the other coefficients $ C_2(4,s) , C_3(4,s) $ and $ C_4(4,s) $. 
\section{Equation and solutions for the equal mass case \labbel{sec:EqMass}}  
\setcounter{equation}{0} 
\numberwithin{equation}{section} 
At the end of Section \ref{sec:DD2} we have shortly considered the equal 
mass limit of the 4th order operator $ D_4(d,s) $, Eq.(\ref{eq:D4}), 
which is still a 4th order operator. In this Section we want to discuss 
the equation for the maxcut of the sunrise amplitude which is obtained 
by starting from scratch in the equal mass case $ m_1 = m_2 = m_3 = m $. 
In that case LiteRed~\cite{Lee2012} gives for the amplitude of the 
maxcut, which will be called $ M(d,s,m^2) $ to emphasize the equality of the 
three masses, the 2nd order equation 
\be D_2(d,s,m^2)\ M(d,s,m^2) = 0 \ , \labbel{eq:2nd} \ee 
where $ D_2(d,s,m^2) $ is the 2nd order differential operator 
\begin{align} D_2(d,s,m^2) =& 
                          \ 2s(s-m^2)(s-9m^2) \frac{d^2}{ds^2}  \nonumber\\ 
 -& \bigl[ 3(d-4)s^2 - 10(d-6)m^2 s - 9d\,m^4 \bigr] \frac{d}{ds} \nonumber\\ 
 +& (d-3)\bigl[ (d-4)s + (d+4)m^2 \bigr]  \ . \labbel{eq:D2} 
\end{align} 
Eq.(\ref{eq:D2}) is to be compared with the corresponding equation valid 
in the different mass case Eq.(\ref{eq:D4M}), which we rewrite here for 
convenience of the reader 
$$ D_4(d,s) M(d,s) = 0 \ . $$ 
We recall once more that in the different mass case the differential 
operator $ D_4(d,s) $ is of 4th order, so that $ D_2(d,s,m^2) $ 
{\it is not} the equal mass limit of $ D_4(d,s) $. 
Note also the (relative) simplicity of $ D_2(d,s,m^2) $ when compared to the 
explicit expression of $ D_4(d,s) $, Eq.s(\ref{eq:D4a}),(\ref{eq:defC4}) 
{\it etc.}, given in the Appendix \ref{app:equ}. 
\par 
On the contrary, the solutions of Eq.(\ref{eq:2nd}) can be derived from the 
functions $ M_i(d,s) $ of Eq.(\ref{eq:Mi}) by considering their equal mass 
limit; we will call them $ M_i(d,s,m^2) $. 
\par 
In the equal mass limit the parameter $ b_1 = (m_1-m_2)^2 $ vanishes, so that 
$ b_1 = 0 $ coincides with $ b_0 = 0 $ and therefore $ M_1(d,s,m^2), $ which 
corresponds to the $ b $ integration in the range from $ b_0 $ to  $ b_1 $, 
is also vanishing. Further, as the very definition of $ R_4(b) $, 
Eq.(\ref{eq:R4}) at $ b_1 = 0 $ gives 
$$ R_4(b) = b(b-b_2)(b-b_3)(b-b_4) \ , $$ 
we introduce 
\be R_3(b) = (b-b_2)(b-b_3)(b-b_4) \ , \labbel{eq:R3} \ee 
so that Eq.(\ref{eq:Mi}) becomes 
\be M_i(d,s,m^2) = s^{\frac{2-d}{2}} \int_{b_{i-1}}^{b_i} db 
    \frac{1}{\sqrt{|b|}} | R_3(b) |^{\frac{d-3}{2}} \ , \labbel{eq:Mim} \ee 
where the index $i$ takes the 5 values (0,2,3,4,5). 
\par 
As a first step, we want to check that the functions $ M_i(d,s,m^2) $ do 
satisfy the 2nd order differential equation Eq.(\ref{eq:2nd}). 
In the different mass case, the differential equation is of 4th order, 
and we needed an expression of the solutions and its 4 derivatives in 
terms of a set of just 4 independent functions, with $ M_i(d,s) $ belonging 
of course to the set. That was achieved by 
means of suitable integration by parts identities ({\it ibp-id}\;'s) for the 
occurring $ b $ integrals, as discussed in the previous Sections. 
Those {\it ibp-id}\;'s, typically, involve 5 terms, and can be used for 
expressing one of the terms as a combination of the other four. 
The presence of 5 terms is related to the presence of 5 factors, namely 
$ b,(b-b_1),(b-b_2),(b-b_3),(b-b_4) $ in the integrands occurring 
in all the {\it ibp-id}\;'s. 
\par 
In the equal mass case, as Eq.(\ref{eq:2nd}) is of 2nd order, one has to 
express each of the $ M_i(d,s,m^2) $ and of its first two $s$-derivatives in 
terms of just two related but independent functions; but the $b$-integrands 
involves 4 monomials (due to $ b_1 = 0 $, $b$ and $ (b-b_1) $ 
coincide), so that (see the continuation for details) 
the typical {\it ibp-id}\;'s involve 4 terms and can be used for expressing 
just one of the four terms as a combination of the other three, leaving us 
with a set of three independent functions, instead of just two independent 
functions, as desired. 
\par 
To start with, let us consider the function 
\be M_3(d,s,m^2) = s^{\frac{2-d}{2}} N_3(d,s,m^2) \ , \labbel{eq:MN3m} \ee 
with 
\be N_3(d,s,m^2) = \int_{b_2}^{b_3} db 
             \frac{1}{\sqrt{b}} R_3(b)^{\frac{d-3}{2}} \ . 
\labbel{eq:N3m} \ee 
As a set of {\it related functions} we can introduce the integrals 
\be N_3^{(k)}(d,s,m^2) = \int_{b_2}^{b_3} db 
     \frac{1}{\sqrt{b}} R_3(b)^{\frac{d-3}{2}} b^k \ , \labbel{eq:N3mk} \ee 
where $ k $ is a positive integer, (with $ N_3(d,s,m^2) $ corresponding 
to $k=0$). We then consider the obvious identities   
\be \int_{b_2}^{b_3} db \frac{d}{db} \biggl[ \frac{1}{\sqrt{b}} 
         R_3(b)^{\frac{d-1}{2}} b^k \biggr] = 0\ , \labbel{eq:IdN3mk} \ee 
valid for $ d > 1 $. By working out the elementary algebra, one obtains for 
each $k$ of Eq.(\ref{eq:IdN3mk}) a 4 terms identity, expressing the integral 
$ N_3^{(k+2)}(d,s,m^2) $ in terms of the three {\it related functions} with 
indices $(k+1), k $ and $ (k-1) $, 
so that all the integrals $ N_3^{(k)}(d,s,m^2) $ can be expressed in terms 
of the three integrals $ N_3(d,s,m^2), N_3^{(1)}(d,s,m^2) $ and 
$ N_3^{(2)}(d,s,m^2) $. 
\par 
We then consider the three {\it auxiliary integrals} 
\be I_3^{(k)}(d,s,m^2) = \int_{b_2}^{b_3} db 
  \frac{1}{\sqrt{b}} R_3(b)^{\frac{d-1}{2}}\ b^k\ , 
  \hspace{4mm} k=0,1,2     \labbel{eq:Ik} \ee 
and their $s$-derivatives 
\be \frac{d}{ds}I_3^{(k)}(d,s,m^2) = \frac{d}{ds}\int_{b_2}^{b_3} db 
  \frac{1}{\sqrt{b}} R_3(b)^{\frac{d-1}{2}}\ b^k\ , 
  \hspace{4mm} k=0,1,2 \ . \labbel{eq:sdIk} \ee 
We can evaluate the $s$-derivatives by following two different paths. 
As a first path, ignoring the $b$-integrals on the {\it r.h.s.} we express 
all the three $ I^{(k)}(d,s,m^2) $ of Eq.(\ref{eq:Ik}) as linear 
combinations, with coefficients depending on $s$, 
of the three {\it related functions} $ N_3^{(k)}(d,s,m^2), k=0,1,2\ $\ by 
using the identities generated by Eq.s(\ref{eq:IdN3mk}), 
and then we carry out the $s$-derivatives of the products of coefficients 
and {\it related functions}; in that way one obtains 
linear combinations of the three $ N_3^{(k)}(d,s,m^2) $ and of their 
$s$-derivatives. Alternatively (second path) we can perform the 
$s$ differentiation of the three $b$ integrands in the {\it r.h.s.} of 
Eq.s(\ref{eq:sdIk}) (for $d>1$ the end-point contributions vanish), 
and then we express the results in terms of the three 
$ N_3^{(k)}(d,s,m^2), k=0,1,2, $ by using again the identities 
Eq.s(\ref{eq:IdN3mk}). As the two paths must obviously give the same results, 
by equating the results one obtains three identities which can be solved 
by expressing the $s$-derivatives of the three {\it related functions} 
$ N_3^{(k)}(d,s,m^2) $ in terms of the same three functions. The result 
can be written as 
\be \frac{d}{ds}N_3^{(k)}(d,s,m^2) = \sum_{j=0,1,2} C_{kj}(d,s) 
                 N_3^{(j)}(d,s,m^2) \ , \hspace{4mm} i=0,1,2 
\labbel{eq:sdN3}  \ee 
with $ N_3^{(0)}(d,s,m^2) = N_3(d,s,m^2) $ as above already specified, and 
\begin{align} 
 C_{00}(d,s) &= - \frac{1}{3s} + \frac{d-2}{s-m^2} 
                - \frac{2}{3(s-9m^2)} \ , \nonumber\\ 
 C_{01}(d,s) &= (3d-4)\biggl[ \frac{1}{4m^2s} - \frac{3}{8m^2(s-m^2)} 
                                                + \frac{1}{8m^2(s-9m^2)} 
                        \biggr] \ , \nonumber\\ 
 C_{02}(d,s) &=  (3d-4)\biggl[ -\frac{1}{12m^4s} + \frac{3}{32m^4(s-m^2) } 
                 - \frac{1}{96m^4(s-9m^2)} \biggr] \ , \labbel{eq:C0k} 
\end{align} 
\begin{align} 
 C_{10}(d,s) &= - \frac{m^2}{3s} - \frac{8m^2}{3(s-9m^2)} \ , \nonumber\\ 
 C_{11}(d,s) &= \frac{3d-4}{4s} - \frac{d}{2(s-m^2)} 
                                + \frac{3d-4}{2(s-9m^2)} \ , \nonumber\\ 
 C_{12}(d,s) &= (3d-4)\biggl[ - \frac{1}{12m^2s} + \frac{1}{8m^2(s-m^2)} 
              - \frac{1}{24m^2(s-9m^2)} \biggr] \ , \labbel{eq:C1k} 
\end{align} 
\begin{align} 
 C_{20}(d,s) &= -m^2 - \frac{m^4}{3s} - \frac{32m^4}{3(s-9m^2)}\ ,\nonumber\\  
 C_{21}(d,s) &= \frac{d}{4} + (3d-4)\frac{m^2}{4s} - \frac{2dm^2}{s-m^2} 
                            + 2\frac{(3d-4)m^2}{s-9m^2} \ , \nonumber\\ 
 C_{22}(d,s) &= (3d-4)\biggl[ -\frac{1}{12s} + \frac{1}{2(s-m^2)} 
              -\frac{1}{6(s-9m^2)} \biggr] \ . \labbel{eq:C2k} 
\end{align} 
The second $s$-derivative of $ N_3(d,s,m^2) $ can then be obtained by 
further differentiating Eq.(\ref{eq:sdN3}) for $ k=0 $, and then by using 
again the Eq.s(\ref{eq:sdN3}),(\ref{eq:C0k}),(\ref{eq:C1k}),(\ref{eq:C2k}).
\par 
Once established the expressions of the $s$-derivative of $ N_3(d,s,m^2) $, 
one can easily obtain the corresponding derivatives of $ M_3(d,s,m^2) $, 
Eq.(\ref{eq:MN3m}), plug them into the 2nd order equation 
Eq.(\ref{eq:2nd}) and check that they do actually satisfy that equation, 
as expected. 
\par 
At first sight, that might look a bit surprising; the fact that 
$ M_3(d,s,m^2) $ and its 
first two derivatives satisfy the 2nd order equation (\ref{eq:2nd}) 
tells us that a relation should exist between those three functions (namely 
$ M_3(d,s,m^2) $ and its two derivatives), but those three functions 
were expressed as combinations of the {\it three} functions 
$ N_3^{(k)}(d,s,m^2),\ k=0,1,2 $ for which we were unable to find 
a relation of linear dependence. 
\par 
A closer investigation shows that by introducing a new 
{\it partner} of $ N_3(d,s,m^2) $, let's call it 
$ \widetilde{N}_3(d,s,m^2), $ defined as 
\begin{equation} 
 \widetilde{N}_3(d,s,m^2) = N_3^{(2)}(d,s,m^2) 
                          - (s+3m^2)N_3^{(1)}(d,s,m^2) \ , \labbel{eq:wtN3} 
\end{equation} 
one can rewrite Eq.(\ref{eq:sdN3}) at $ k=0 $ as 
\begin{align} 
  \frac{d}{ds}N_3(d,s,m^2) &= 
  \left( - \frac{1}{3s} + \frac{d-2}{s-m^2} - \frac{2}{3(s-9m^2)} \right) 
                                                  N_3(d,s,m^2) \nonumber\\ 
  &+ \frac{3d-4}{m^4} \left( - \frac{1}{12s} + \frac{3}{32(s-m^2)} 
   - \frac{1}{96(s-9m^2)} \right) \widetilde{N}_3(d,s,m^2) \ . 
                        \labbel{eq:sdN3a} 
\end{align} 
One further finds that the derivative of $ \widetilde{N}_3(d,s,m^2) $ 
can be written as 
\begin{align} 
  \frac{d}{ds}\widetilde{N}_3(d,s,m^2) &= 
  \left( 2m^2 + \frac{2m^4}{3s} + \frac{64m^4}{3(s-9m^2)} \right) 
                                                  N_3(d,s,m^2) \nonumber\\ 
  &+ (3d-4) \left( \frac{1}{6s} + \frac{1}{3(s-9m^2)} \right) 
                       \widetilde{N}_3(d,s,m^2) \ , \labbel{eq:sdwtN3} 
\end{align} 
{\it i.e.} can also be expressed in terms of the same two functions 
$ N_3(d,s,m^2) $ and $ \widetilde{N}_3(d,s,m^2) $. \par 
By repeated use of Eq.s(\ref{eq:sdN3a},\ref{eq:sdwtN3}) one can then 
express all the $s$-derivatives of $ N_3(d,s,m^2) $ 
and its {\it partner} $ \widetilde{N}_3(d,s,m^2) $ in terms of the same 
two functions and then verify that 
$$ M_3(d,s,m^2) = s^\frac{2-d}{2}N_3(d,s,m^2) $$ satisfies the second order 
differential equation Eq.(\ref{eq:2nd}) introduced at the beginning 
of this Section. 
\par 
Besides Eq.(\ref{eq:wtN3}), let us consider the function 
\be S_3(d,s) = N_3^{(1)}(d,s,m^2) - \frac{1}{3}(s+3m^2)N_3^{(0)}(d,s,m^2) \ ; 
            \labbel{eq:Sab} \ee 
remarkably, thanks to Eq.s(\ref{eq:sdN3}), its $s$-derivative is given 
by the simple expression 
\be \frac{d}{ds}S_3(d,s) = \frac{d-2}{s-m^2}S_3(d,s) \ . \labbel{eq:sdSab} \ee 
\par 
Eq.(\ref{eq:sdSab}) is actually a 1st order differential equation in $s$, 
which can be immediately solved giving 
\be S_3(d,s) = X_3(d)\ (s-m^2)^{d-2} \ , \labbel{eq:S3} \ee 
where $ X_3(d) $ is an as yet unspecified normalization constant 
(later on, we will give its value for d =2,3 ). 
\par 
It is easy to check that 
the arguments so far used for deriving the properties of $ N_3(d,s,m^2) $ 
of Eq.(\ref{eq:N3m}) rely on the absence of end-point contributions 
in the various {\it ibp-id}\;'s, and therefore apply as well to 
\be N_2(d,s,m^2) = \int_{b_1=0}^{b_2} db 
             \frac{1}{\sqrt{b}} R_3(b)^{\frac{d-3}{2}} \ . 
\labbel{eq:N2m} \ee 
and 
\be N_4(d,s,m^2) = \int_{b_3}^{b_4} db 
             \frac{1}{\sqrt{b}} R_3(b)^{\frac{d-3}{2}} \ ; 
\labbel{eq:N4m} \ee 
further, they apply also to 
\be N_0(d,s,m^2) = \int_{0}^{\infty} dc 
             \frac{1}{\sqrt{c}} R_3(c)^{\frac{d-3}{2}} 
\labbel{eq:N0m} \ee 
(see Eq.(\ref{eq:N0}) for the notation); 
indeed, the convergence at $ c=0 $ is independent of $ d $, so that the 
convergence at large $ c $ can be achieved for $ d $ sufficiently small 
(in fact negative), and the results can be extended by analytic 
continuation also to positive values of $ d $. 
\par 
We can then generalize Eq.s(\ref{eq:Sab},\ref{eq:sdSab},\ref{eq:S3}) into 
\begin{align} 
  S_i(d,s) &= N_i^{(1)}(d,s,m^2) - \frac{1}{3}(s+3m^2)N_i^{(0)}(d,s,m^2) \ , 
            \labbel{eq:Sabi} \\ 
  \frac{d}{ds}S_i(d,s) &= \frac{d-2}{s-m^2}S_i(d,s) \ , \labbel{eq:sdSabi} 
\end{align} 
valid for $ i=0,2,3,4 $. The above equation is easily integrated, giving 
the obvious generalization of Eq.(\ref{eq:S3}) 
\be S_i(d,s) = X_i(d)\ (s-m^2)^{(d-2)} \ , \labbel{eq:inSabi} \ee 
where the constants $ X(d,i) $ are still to be fixed. 
\par 
Let us also note that, by defining 
\be T_i^{(n)}(d,s) = (s-m^2)^n S_i(d,s) \ , \labbel{eq:Ti}     \ee 
Eq.(\ref{eq:sdSabi} ) becomes 
\be 
  \frac{d}{ds}T_i^{(n)}(d,s) = \frac{d-n-2}{s-m^2}T_i^{(n)}(d,s) \ , 
\labbel{eq:sdTi} \ee 
giving in particular 
\be 
  \frac{d}{ds}T_i^{(2)}(d,s) = \frac{d-4}{s-m^2}T_i^{(2)}(d,s) \ . 
\labbel{eq:sdT2} \ee 
\par 
The case of 
\be N_5(d,s,m^2) = \int_{b_4}^{\infty} db 
             \frac{1}{\sqrt{b}} R_3(b)^{\frac{d-3}{2}} \ ; 
\labbel{eq:N5m} \ee 
is different; if the {\it related functions} are defined as in 
Eq.(\ref{eq:N3mk}), the convergence of the factor 
$ R_3(b)^{\frac{d-3}{2}} $ at $ b = b_4 = (E+m)^2 $ requires $ d > 1 $, 
which is incompatible with the convergence of all the same 
{\it related functions} at large $ b $. 
\par 
In this case one can however introduce as {\it related functions} the 
integrals 
\be N_5^{(k)}(d,s,m^2) = \int_{b_4}^{\infty} db 
     \frac{1}{\sqrt{b}} R_3(b)^{\frac{d-3}{2}} \frac{1}{b^k} \ , 
\labbel{eq:N5mk} \ee 
with $ k \ge 0 $, and as {\it auxiliary} integrals 
\be I_5^{(k)}(d,s,m^2) = \int_{b_4}^{\infty} db 
  \frac{1}{\sqrt{b}} R_3(b)^{\frac{d-1}{2}}\ \frac{1}{b^k}\ , 
  \hspace{4mm} k=0,1,2 \ .    \labbel{eq:I5k} \ee 
It is then found that by introducing as {\it partner} function 
\begin{align} 
    \widetilde{N}_5(d,s,m^2) &= 3m^2(s-m^2)^2(s+3m^2) N_5^{(2)}(d,s,m^2) 
                                                     \nonumber\\ 
     &+ \left( \frac{d-4}{4}(s+3m^2)^3 + 3m^2(s-m^2)^2 \right) 
                             N_5^{(1)}(d,s,m^2 ) \ , \labbel{eq:wtN5} 
\end{align} 
one can verify that all the $s$-derivatives of $ N_5(d,s,m^2) $ and 
$ \widetilde{N}_5(d,s,m^2) $, in particular the 1st and the 2nd 
derivative of $ N_5(d,s,m^2) $, can be expressed in terms of the two 
same functions $ N_5(d,s,m^2), \widetilde{N}_5(d,s,m^2) $, and, finally, 
that also $ N_5(d,s,m^2) $ does satisfy Eq.(\ref{eq:2nd}), as expected. 
We were however unable to find the analog of Eq.(\ref{eq:sdSabi}) for 
$ N_5(d,s,m^2) $. 
\par 
Coming back to Eq.(\ref{eq:Sabi}), skipping the case of $ S_0(d,s) $ which 
would require an analytical continuation for $ d = 3,4 $, 
let us observe that for $ i=2,3,4 $ it can be written as 
\be S_i(d,s) = \int_{b_{i-1}}^{b_i} \frac{db}{\sqrt{b}} 
   |R_3(b,m^2)|^{\frac{d-3}{2}|} \left( b - \frac{1}{3}(s+3m^2) \right) \ , 
\labbel{eq:Sabid} \ee 
which for $ d=2 $ Eq.(\ref{eq:Sabi}) becomes 
\be S_i(2,s) = \int_{b_{i-1}}^{b_i} \frac{db} 
             {|\sqrt{b(b-4m^2}\sqrt{R_2(s,b,m^2)}|} 
             \left( b - \frac{1}{3}(s+3m^2) \right) \ , 
\labbel{eq:Sabi2} \ee 
with $ R_2(s,b,m^2) $ given, as usual, from Eq.(\ref{eq:R2m123}). 
\par 
It can be recalled here the formula (see~\cite{Barbieri:1974nc}, after 
fixing a typo)  
\be \frac{d}{db}\ln{\left( 
        \frac{b(s+3m^2-b)+\sqrt{b(b-4m^2)}\sqrt{R_2(s,b,m^2)}} 
             {b(s+3m^2-b)-\sqrt{b(b-4m^2)}\sqrt{R_2(s,b,m^2)}} \right)} 
   = \frac{s+3m^2-3b}{\sqrt{b(b-4m^2)}\sqrt{R_2(s,b,m^2)}} \ , 
   \labbel{eq:Sablog} \ee 
whose integration in $ b $ gives at once the formula (exact) 
\be S_i(2,s) = - \frac{1}{3} \ln{\left( 
   \frac{b(s+3m^2-b)+\sqrt{b(b-4m^2)}\sqrt{R_2(s,b,m^2)}} 
        {b(s+3m^2-b)-\sqrt{b(b-4m^2)}\sqrt{R_2(s,b,m^2)}} \right)} 
    \Biggr|_{b_{i-1}}^{b_i} \ ,        \labbel{eq:Sabilg} 
\ee 
where the proper analytical continuation of the logarithm is understood. 
The above equation gives 
\begin{align}  S_2(2,s) &= - \frac{1}{3}\pi \ , \nonumber\\ 
               S_3(2,s) &= 0 \ ,                \nonumber\\ 
               S_4(2,s) &= + \frac{2}{3}\pi \ . \labbel{eq:Sabi2c} 
\end{align} 
To our knowledge, for $ i = 3 $ Eq.s(\ref{eq:Sabi2},\ref{eq:Sabi2c}) 
reproduce a result contained in equation (85) of the Sabry paper~\cite{Sabry}, 
(see Eq.s(32,67,68) of that paper for the notations). 
\\ 
For $ d=3,5 $ the integration of Eq.(\ref{eq:Sabid}) is trivial; one finds 
\begin{align}  S_2(3,s) &= - \frac{4}{3}m(s-m^2) \ , \nonumber\\ 
               S_3(3,s) &= 0 \ , \nonumber\\ 
               S_4(3,s) &= + \frac{8}{3}m(s-m^2) \ , \labbel{eq:Sabi3c} 
\end{align} 
and 
\begin{align}  S_2(5,s) &= - \frac{32}{9}m^3(s-m^2)^3 \ , \nonumber\\ 
               S_3(5,s) &= 0 \ , \nonumber\\ 
               S_4(5,s) &= + \frac{64}{9}m^3(s-m^2)^3 \ . \labbel{eq:Sabi5c} 
\end{align} 
For $ d=4 $ the direct calculation is not possible; we used the large $ s $ 
expansion for $ S_2(4,s), S_4(4,s) $, and an expansion just avove the 
threshold $ s = (3m)^2 $ (we recall that our integration formulas 
are valid for $ (3m)^2 < s < +\infty $). We find 
\begin{align}  S_2(4,s) &= - \frac{2}{3}\pi\ m^2(s-m^2)^2 \ , \nonumber\\ 
               S_3(4,s) &= 0 \ , \nonumber\\ 
               S_4(4,s) &= + \frac{4}{3}\pi\ m^2(s-m^2)^2 \ . 
\labbel{eq:Sabi4c} 
\end{align} 
The above equations are (of course) in agreent with Eq.(\ref{eq:S3}). 
\section*{Acknowledgements} 
We are grateful to J. Vermaseren for his assistance in the use of the 
algebraic program FORM~\cite{Vermaseren:2000nd} which was intensively used 
in all the steps of this work. 
\\ \noindent
\appendix 
\section{Evaluation of the Maxcut} \labbel{app:max} 
\setcounter{equation}{0} 
\numberwithin{equation}{section} 
We give some details of the evaluation of the integral appearing in Section 
\ref{sec:Mc}, starting from $ B(d,-k^2) $ Eq.(\ref{eq:Bk2}), 
$$ B(d,-k^2) = \int \D^dq\ \delta(D_1)\delta(D_2) \ . $$ 
As a first step, due to the presence of the factor $ \delta(D_1) $ in the 
integrand, one can write 
\be B(d,-k^2) = \int \D^dq\ \delta(D_1)\delta(D_2-D_1) \ , \labbel{eq:B1} \ee 
and from Eq.s(\ref{eq:D123}) we have at once 
$$ D_2 - D_1 = k^2 - 2(k\cdot q) - m_1^2 + m_2^2 \ . $$ 
According to Eq.(\ref{eq:k2}) (where the Minkoskean regularisation is used) 
we then write $ (k\cdot q) $ as 
$$ (k\cdot q) = -k_0q_0 + k_xq_x - (\vec{K}\cdot \vec{Q}) 
              = -k_0q_0 + k_xq_x - K Q_1 \ , $$ 
where K is the modulus of $\vec{K}$, while the first component of 
$ \vec{Q} $, namely $Q_1$, is the component of $\vec{Q}$ along the direction 
of $\vec{K}$. By performing the change of integration variables 
$ (q_0,Q_1) \to ({q_0}',{Q_1}') $ 
(which is in fact a rotation, with Jacobian equal 1) 
\begin{align} 
   q_0 &= \frac{1}{\sqrt{k_0^2+K^2}}(k_0 {q_0}' - K   {Q_1}') \ ,\nonumber\\ 
   Q_1 &= \frac{1}{\sqrt{k_0^2+K^2}}(  K {q_0}' + k_0 {Q_1}') \ ,\nonumber 
\end{align} 
and then by rewriting again $ (q_0,Q_1) $ instead of $ ({q_0}',{Q_1}') $, 
one finds 
$$ (k\cdot q) = - \sqrt{k_0^2+K^2}\ q_0 + k_xq_x \ , $$ 
where $ Q_1 $ has disappeared. At this point one has 
\begin{align} 
  D_1 &= - q_0^2 + q_x^2 - Q^2 + m_1^2 \ , \labbel{eq:D1}\\ 
  D_2 - D_1 &= 2\sqrt{k_0^2+K^2}\ q_0  - 2k_xq_x + k^2 - m_1^2 + m_2^2 \ , 
                                           \labbel{eq:D21}  
\end{align} 
and, as the components of $ \vec{Q} $ do not appear explicitly in the 
integrand, we can introduce polar coordinates in $ \D^dq $, Eq.(\ref{eq:Ddq}), 
$$ \D^dq = dq_0\ dq_x\ d^{d-2}Q 
         = dq_0\ dq_x\ \Omega(d-2) \int_0^\infty Q^{d-3}\ dQ \ , $$ 
where 
\be \Omega(n) = 2 \frac{\pi^{\frac{n}{2}}} 
                       {\Gamma\left(\frac{n}{2}\right)} \ , \labbel{eq:Om} \ee 
is the solid angle in $n$ dimensions, so that Eq.(\ref{eq:B1}) becomes 
\be   B(d,-k^2) = \Omega(d-2) \int dq_0\ dq_x 
      \int_0^\infty Q^{d-3}\ dQ \ \delta(D_1)\delta(D_2-D_1) \ . 
\labbel{eq:B2} \ee 
\par 
Let us consider separately the cases of timelike and spacelike $k$. \par 
If $k$ is timelike, ($ -k^2 = k_0^2-k_x^2+K^2 > 0$), we perform the 
change of variable (a Lorentz boost, with Jacobian equal 1) 
\begin{align} 
   q_0 &= \frac{1}{\sqrt{-k^2}} \left( 
                \sqrt{k_0^2+K^2} {q_0}' + k_x{q_x}' \right) \ ,\nonumber\\ 
   q_x &= \frac{1}{\sqrt{-k^2}} \left( 
                k_x {q_0}' + \sqrt{k_0^2+K^2}{q_x}' \right) \ ,\nonumber 
\end{align} 
and then we rewrite again $ (q_0,q_x) $ instead of $ ({q_0}',{q_x}'). $ 
Eq.(\ref{eq:D1}) does not change, while Eq.(\ref{eq:D21}) becomes 
\be D_2 - D_1 = 2\sqrt{-k^2}\ q_0 + k^2 - m_1^2 + m_2^2 \ . 
                                            \labbel{eq:D21a} \ee 
By integrating $ \delta(D_2-D_1) $ in $ q_0 $ and $ \delta(D_1) $ in $ q_x $ 
Eq.(\ref{eq:B2}) becomes (for timelike $k$) 
\be \Theta(-k^2)B(d,-k^2) = {\Theta}(-k^2)\ \Omega(d-2) 
   \frac{1}{2\sqrt{-k^2}} \int_0^\infty dQ 
   \frac{Q^{d-3} } {\sqrt{Q^2-\frac{1}{4k^2}R_2(-k^2,m_1^2,m_2^2)}} \ , 
\labbel{eq:B2t} \ee 
where 
\be R_2(a,b,c) = a^2 + b^2 + c^2 - 2ab - 2ac - 2bc \ , \labbel{eq:R2abc} 
\ee 
in agreement with Eq.(\ref{eq:R2m123}), 
and it is understood that the argument of the square root must be positive, 
{\ie} the constraint 
\be Q^2-\frac{1}{4k^2}R_2(-k^2,m_1^2,m_2^2) \ge 0 \ \labbel{eq:Q2c} \ee 
must hold. 
\par 
If $ R_2(-k^2,m_1^2,m_2^2) $ is positive, as $ k^2 $ is negative, 
Eq.(\ref{eq:Q2c}) is always satisfied; one can rescale $ Q^2 \to v $ 
according to 
$$ Q^2 = - \frac{1}{4k^2}R_2(-k^2,m_1^2,m_2^2)\ v \ , $$ 
and the calculation is completed with the integral 
$$  \int_0^\infty dv\ v^{\frac{d-4}{2}}\ (1+v)^{-\frac{1}{2}} = 
  B\left(\frac{d-2}{2},\frac{3-d}{2}\right) 
  = \frac{1}{\sqrt{\pi}}\Gamma\left(\frac{d-2}{2}\right) 
                 \Gamma\left(\frac{3-d}{2}\right) \ , $$ 
where $ B(x,y), \Gamma(x) $ are the Euler's Beta and Gamma functions. \par 
If $ R_2(-k^2,m_1^2,m_2^2) $ is negative, as $ k^2 $ is also negative, 
Eq.(\ref{eq:Q2c}) gives the condition 
$$ Q^2 \ge \frac{1}{4k^2}R_2(-k^2,m_1^2,m_2^2) \ ; $$ 
one can then rescale 
$$ Q^2 = + \frac{1}{4k^2}R_2(-k^2,m_1^2,m_2^2)\ w \ , $$ 
and the calculation is completed with the integral 
$$  \int_1^\infty dw\ w^{\frac{d-4}{2}}\ (w-1)^{-\frac{1}{2}} = 
   B\left(\frac{3-d}{2},\frac{1}{2}\right) = 
   \frac{\sqrt{\pi}}{\Gamma\left(\frac{4-d}{2}\right)}
   \Gamma\left( \frac{d-3}{2}\right) \ . $$ 
\par 
If $k$ is spacelike, ($ k^2 = - k_0^2+k_x^2-K^2 > 0$), in analogy with the 
procedure followed for $k$ timeike, we perform the 
change of variable (again a Lorentz boost, with Jacobian equal 1) 
\begin{align} 
   q_0 &= \frac{1}{\sqrt{k^2}} \left( 
                k_x {q_0}' + \sqrt{k_0^2+K^2} {q_x}' \right) \ ,\nonumber\\ 
   q_x &= \frac{1}{\sqrt{k^2}} \left( 
                \sqrt{k_0^2+K^2} {q_0}' + k_x {q_x}' \right) \ ,\nonumber 
\end{align} 
we rewrite again $ (q_0,q_x) $ instead of $ ({q_0}',{q_x}'), $ 
and then we integrate $ \delta(D_2-D_1) $ in $ q_x $ and $ \delta(D_1) $ 
in $ q_0 $; Eq.(\ref{eq:B2}) becomes (for spacelike $k$) 
\be \Theta(k^2)B(d,-k^2) = {\Theta}(k^2)\ \Omega(d-2) 
   \frac{1}{2\sqrt{k^2}} \int_0^\infty dQ 
   \frac{Q^{d-3} } {\sqrt{\frac{1}{4k^2}R_2(-k^2,m_1^2,m_2^2)-Q^2}} \ , 
\labbel{eq:B2s} \ee 
and it is understood that the argument of the square root must be positive, 
{\ie} the constraint 
\be \frac{1}{4k^2}R_2(-k^2,m_1^2,m_2^2) - Q^2 \ge 0 \ \labbel{eq:Q2d} \ee 
must hold. Note that we can write 
$$ R_2(-k^2,m_1^2,m_2^2) = (k^2+(m_1-m_2)^2)(k^2+(m_1+m_2)^2) \ , $$ 
so that, for $ k^2 $ positive ($ k $ spacelike) $ R_2(-k^2,m_1^2,m_2^2) $ 
is always positive; we can then rescale $ Q^2 \to t $ according to 
$$ Q^2 = + \frac{1}{4k^2}R_2(-k^2,m_1^2,m_2^2)\ t \ , $$ 
and the calculation is completed with the integral 
$$ \int_0^1 dt\ t^{\frac{d-4}{2}}\ (1-t)^{-\frac{1}{2}} = 
   B\left( \frac{d-2}{2}, \frac{1}{2} \right) = 
   \frac{\sqrt{\pi}}{\Gamma\left(\frac{d-1}{2}\right)} 
   \Gamma\left( \frac{d-2}{2}\right) \ . $$ 
\par 
By collecting results we have 
\begin{align} B(d,-k^2) = 
  & \ \Theta(-k^2)\Theta(+R_2(-k^2,m_1^2,m_2^2) ) 
         \ |k^2|^{\frac{2-d}{2}}\ |R_2(-k^2,m_1^2,m_2^2)|^{\frac{d-3}{2}} 
  \nonumber\\ 
  & {\hskip3mm} \times 2^{1-d} \Omega(d-2) 
    \ \frac{1}{\sqrt{\pi}}\Gamma\left( \frac{d-2}{2}\right) 
                          \Gamma\left(\frac{3-d}{2}\right) \nonumber\\ 
  + & \ \Theta(-k^2)\Theta(-R_2(-k^2,m_1^2,m_2^2) ) 
         \ |k^2|^{\frac{2-d}{2}}\ |R_2(-k^2,m_1^2,m_2^2)|^{\frac{d-3}{2}} 
  \nonumber\\ 
  & {\hskip3mm} \times 2^{1-d} \Omega(d-2) 
    \ \frac{\sqrt{\pi}}{\Gamma\left(\frac{4-d}{2}\right)} 
                \Gamma\left(\frac{3-d}{2}\right) \nonumber\\ 
%
%
%
  + & \ \Theta(+k^2)\Theta(+R_2(-k^2,m_1^2,m_2^2) ) 
         \ |k^2|^{\frac{2-d}{2}}\ |R_2(-k^2,m_1^2,m_2^2)|^{\frac{d-3}{2}} 
  \nonumber\\ 
  & {\hskip3mm} \times 2^{1-d} \Omega(d-2) 
    \ \frac{\sqrt{\pi}}{\Gamma\left(\frac{d-1}{2}\right)} 
                        \Gamma\left( \frac{d-2}{2}\right) \ . 
\labbel{eq:B3} 
\end{align} 
The evaluation of $ C(d,b) $, Eq.(\ref{eq:C}) is similar, with $ p^2 $ 
playing the role of $ k^2 $ in $ B(d,-k^2) $. 
For simplicity, we restrict ourself to consider only the case of timelike 
$ p $, and above the unitarity threshold, \ie 
$$ s = -p^2 = E^2 \ , {\hskip5mm} E > (m_1+m_2+m_3) \ , $$ 
we find 
\begin{align} C(d,E^2,b) = 
  & \ \Theta(+R_2(E^2,b,m_3^2) ) 
         \ E^{2-d}\ |R_2(E^2,b,m_3^2)|^{\frac{d-3}{2}} 
  \nonumber\\ 
  & {\hskip3mm} \times 2^{1-d} \Omega(d-2) 
    \ \frac{1}{\sqrt{\pi}}\Gamma\left(\frac{d-2}{2}\right) 
                          \Gamma\left(\frac{3-d}{2}\right) \nonumber\\ 
  + & \ \Theta(-R_2(E^2,b,m_3^2) ) 
         \ E^{2-d}\ |R_2(E^2,b,m_3^2)|^{\frac{d-3}{2}} 
  \nonumber\\ 
  & {\hskip3mm} \times 2^{1-d} \Omega(d-2) 
    \ \frac{\sqrt{\pi}}{\Gamma\left(\frac{4-d}{2}\right)} 
                \Gamma\left(\frac{3-d}{2}\right) \ . \labbel{eq:C1} 
\end{align} 
The integrand in $b$ of $ M(d,s) $, Eq.(\ref{eq:Mb}), is the product 
of $ B(d,b) $ and $ C(d,E^2,b) $, Eq.s(\ref{eq:B3},\ref{eq:C1}). 
From the definition of $ R_2(a,b,c) $, Eq.(\ref{eq:R2abc}), one finds 
\begin{align} 
  R_2(b,m_1^2,m_2^2) &= (b-(m_1-m_2)^2) (b-(m_1+m_2)^2) 
                      = (b-b_1)(b-b_2) \ , \nonumber\\ 
  R_2(E^2,b,m_3^2)   &= (b-(E-m_3)^2) (b-(E+m_3)^2) 
                      = (b-b_3)(b-b_4) \ , \nonumber\\ 
\end{align} 
where the $ b_i $ are defined as in Eq.s(\ref{eq:bi}), so that $ R_4(b) $, 
Eq.(\ref{eq:R4}), is the product of $ R_2(b,m_1^2,m_2^2) $ and 
$ R_2(E^2,b,m_3^2) $, 
$$ R_4(b) = R_2(b,m_1^2,m_2^2)\ R_2(E^2,b,m_3^2) \ . $$ 
The product of $ B(d,b) $ and $ C(d,E^2,b) $ contains also the product of 
several $\Theta$-functions whose arguments depend on $ b $, see the 
formulas above. To work out the constraints implied by those $ \Theta $'s, 
let us introduce a new $\Theta$-function with 3 arguments, 
$\Theta(b_i,b,b_j) $, where $ b_i, b_j$ are defined as in Eq.s(\ref{eq:bi}) 
and $ b_i < b_j $ (an essential condition), whose value is 1 for $b$ in the 
range $ b_i < b < b_j $, and zero otherwise. We have, obviously 
\begin{align} 
 &\Theta(b) = \Theta(b_0,b,b_5) \ , {\hskip3mm} 
  \Theta(-b) = \Theta(b_{-1},b,b_0) \ , \nonumber\\ 
 &\Theta(+R_2(b,m_1^2,m_2^2)) = \Theta(b_{-1},b,b_1) + \Theta(b_2,b,b_5) \ , 
  {\hskip3mm} 
  \Theta(-R_2(b,m_1^2,m_2^2)) = \Theta(b_1,b,b_2) \ , \nonumber\\ 
 &\Theta(+R_2(E^2,b,m_3^2)) = \Theta(b_{-1},b,b_3) + \Theta(b_4,b,b_5) \ , 
  {\hskip3mm} 
  \Theta(-R_2(E^2,b,m_3^2)) = \Theta(b_3,b,b_4) \ . \nonumber 
\end{align} 
As a next step, consider separately each term $\Theta(b_i,b,b_j) $; if 
$ b_i $ and $b_j $ are not adjacent, {\ie} if $ i \ne j-1 $, write it as 
a sum of adjacent terms, as in the following examples 
\begin{align} 
 \Theta(b_{-1},b,b_3) &= \Theta(b_{-1},b,b_0) + \Theta(b_0,b,b_1) 
                      + \Theta(b_1,b,b_2) + \Theta(b_2,b,b_3) \ , \nonumber\\ 
 \Theta(b_2,b,b_5) &= \Theta(b_2,b,b_3) + \Theta(b_3,b,b_4) 
                      + \Theta(b_3,b,b_4) \ . \nonumber 
\end{align} 
In so doing one obtains a proliferation of terms, consisting of the products 
of several $ \Theta $-functions with 3 arguments; as a last step, use the 
(obvious) identity 
$$ \Theta(b_{i-1},b,b_i)\ \Theta(b_{j-1},b,b_j) = \delta_{i,j} 
                                              \ \Theta(b_{i-1},b,b_i) $$ 
(the products of two $\Theta$'s is zero unless the arguments are identical) 
and one is left with few terms containing a single $\Theta$ factor. 
By observing that 
$$ \int_{-\infty}^{\infty} db\ \Theta(b_{i-1},b,b_i) \ f(b) 
       = \int_{b_{i-1}}^{b_i} db\ f(b) \ , $$ 
Eq.s(\ref{eq:Mc},\ref{eq:Mi}) are recovered. 
For completeness, we report here the explicit values of the coefficients 
$ C_i(d) $ appearing in Eq.(\ref{eq:Mc}) (even if they don't play any role 
in this paper) 
\begin{align} 
  C_0(d) &=  2^{2d-3} \Omega(d-1)\frac{1}{\sqrt{\pi}} 
                      \Gamma\left( \frac{d-2}{2} \right) 
                      \Gamma\left( \frac{3-d}{2} \right) \ , \nonumber\\ 
  C_1(d) &=  2^{2d-3} \Omega(d-1)\frac{1}{\pi\sqrt{\pi}} 
                      \Gamma\left( \frac{d-1}{2} \right) 
                      \Gamma\left( \frac{d-2}{2} \right) 
                    \Gamma^2\left( \frac{3-d}{2} \right) \ , \nonumber\\ 
  C_2(d) &=  2^{2d-3} \Omega(d-1)\frac{1}{\sqrt{\pi} 
                      \Gamma\left( \frac{4-d}{2} \right)} 
                      \Gamma\left( \frac{d-1}{2} \right) 
                    \Gamma^2\left( \frac{3-d}{2} \right) \ , \nonumber\\ 
  C_3(d) &=  2^{2d-3} \Omega(d-1)\frac{1}{\pi\sqrt{\pi}} 
                      \Gamma\left( \frac{d-1}{2} \right)
                      \Gamma\left( \frac{d-2}{2} \right)
                    \Gamma^2\left( \frac{3-d}{2} \right) \ , \nonumber\\ 
  C_4(d) &=  2^{2d-3} \Omega(d-1)\frac{1}{\sqrt{\pi} 
                      \Gamma\left( \frac{4-d}{2} \right)} 
                    \Gamma^2\left( \frac{3-d}{2} \right) \ , \nonumber\\ 
  C_5(d) &=  2^{2d-3} \Omega(d-1)\frac{1}{\pi\sqrt{\pi}} 
                      \Gamma\left( \frac{d-1}{2} \right) 
                      \Gamma\left( \frac{d-2}{2} \right) 
                    \Gamma^2\left( \frac{3-d}{2} \right) \ . \labbel{eq:defCi} 
\end{align} 
\section{The Differential Equations} \labbel{app:equ} 
\setcounter{equation}{0} 
\numberwithin{equation}{section}  
The $4$th order differential equation for the {\it maxcut} of the Sunrise 
amplitude was generated by the Mathematica package LiteRed~\cite{Lee2012}.
The equation was already written in Eq.(\ref{eq:D4M}), which we repeat here 
for the convenience of the reader 
\begin{equation}
    D_4(d,s,m_i^2)\ M(d,s) = 0 \ , \labbel{eq:4th} 
\end{equation}
where $ M(d,s) $ is a generic solution, and $ D_4(d,s,m_i^2) $ is the 4th 
order differential operator, already written in Eq.(\ref{eq:D4a}) and 
again rewritten here 
\begin{align} 
 D_4(d,s,m_i^2) &= C_4(d,s,m_i^2) \frac{d^4}{ds^4} 
                 + C_3(d,s,m_i^2) \frac{d^3}{ds^3} \nonumber\\ 
                &+ C_2(d,s,m_i^2) \frac{d^2}{ds^2} 
                 + C_1(d,s,m_i^2) \frac{d}{ds} 
                 + C_0(d,s,m_i^2)\ , \labbel{eq:D4a} 
\end{align} 
The 5 coefficients $ C_i(d,s,m_i^2), i=0,1,..,4 $ are 
polynomials in $d, s$ and the three masses $m_i, i=1,2,3$. 
As the equation and all those polynomials are symmetrical under the 
permutations of the three masses, it can be convenient to express them 
through the three symmetrical combinations of the three masses, namely 
\begin{align} 
 U =&\ m_1^2 + m_2^2 + m_3^2                     \ ,    \nonumber\\ 
 D =&\ m_1^2 m_2^2 + m_2^2 m_3^2 + m_3^2 m_1^2   \ ,    \nonumber\\ 
 T =&\ m_1^2 m_2^2 m_3^2 \ . \labbel{eq:UDT} 
\end{align} 
With the above definition, $ R_2(m_1^2,m_2^2,m_3^2) $, defined as in 
Eq.(\ref{eq:R2abc}), becomes 
\be R_2(m_1^2,m_2^2,m_3^2) = U^2 - 4 D \ . \labbel{eq:R2UD} \ee 
\par 
The explicit expression of the coefficient $ C_4(d,s,m_i^2) $ is 
\begin{equation}
  C_4(d,s,m_i^2) = 8s^3 \ \mathcal{D}(s) \ 
  \bigg[ (7-d)s^2 + 2(3-d)s\ U + (3d-13)(U^2-4D) \bigg] 
  \labbel{eq:defC4} 
\end{equation}
where 
\begin{align}
    \mathcal{D}(s) = 
   &\ (s-(m_1+m_2+m_3)^2)(s-(m_1+m_2-m_3)^2)       \nonumber\\ 
   &\ (s-(m_1-m_2+m_3)^2)(s-(m_1-m_2-m_3)^2)       \nonumber\\ 
  =&\ s^4 - 4 s^3 U + s^2(6U^2 - 8D) + s(16 UD - 4 U^3 - 64 T) 
          + (U^2 - 4 D)^2 \ . 
\labbel{eq:DB3} 
\end{align} 
We further have 
\begin{align} 
 C_3(d,s,m_i^2) = 
    &\ s^8 ( 1008 - 284 d + 20 d^2 ) 
     + s^7 ( - 2000 + 328 d - 8 d^2 ) U                      \nonumber\\ 
   +&\ s^6 \big[\ ( 4736 - 2224 d + 208 d^2 )D 
                + (  - 2080 + 1372 d - 132 d^2 )U^2 \ \big] \nonumber\\ 
   +&\ s^5 \big[\ (  - 10752 - 256 d + 256 d^2 ) T 
                + ( 8160 - 3088 d + 208 d^2 ) U^3            \nonumber\\ 
    & \hspace{8mm} 
                + ( - 25728 + 9664 d - 704 d^2 ) UD \ \big] \nonumber\\ 
   +&\ s^4 \big[\ ( - 35072 + 10944 d - 576 d^2 ) D^2 
                + ( - 7120 + 2012 d - 52 d^2 ) U^4           \nonumber\\ 
    & \hspace{8mm} 
                + ( 37248 - 10784 d + 352 d^2 ) U^2 D 
                + ( - 12288 + 2560 d + 512 d^2 ) U T \ \big] \nonumber\\ 
   +&\ s^3 \big[\ ( - 133120 + 17408 d + 3072 d^2 ) D T 
                + ( 2032 - 184 d - 72 d^2 ) U^5              \nonumber\\ 
    & \hspace{8mm} 
                + ( - 16256 + 1472 d + 576 d^2 ) U^3 D 
                + ( 33280 - 4352 d - 768 d^2 ) U^2 T         \nonumber\\ 
    & \hspace{8mm} 
                + ( 32512 - 2944 d - 1152 d^2 ) U D^2 \ \big] \nonumber\\ 
   +&\ s^2 \big[\ ( 9984 d - 2304 d^2 )D^3 
                + ( - 156 d + 36 d^2 )U^6                   \nonumber\\ 
    & \hspace{8mm} 
                + ( 1872 d - 432 d^2 ) U^4 D 
                + ( - 7488 d + 1728 d^2 ) U^2 D^2 \ \big]  \ , 
\labbel{eq:defC3} 
\end{align} 
\begin{align} 
 C_2(d,s,m_i^2) = 
     &\ s^7 ( 5376 - 2448 d + 366 d^2 - 18 d^3 ) 
      + s^6 ( - 4720 + 652 d + 124 d^2 - 16 d^3 ) U              \nonumber\\ 
    +&\ s^5 \big[\ ( 44832 - 24528 d + 4152 d^2 - 216 d^3 ) D 
                + (  - 18720 + 11436 d - 1998 d^2 + 102 d^3 ) U^2 
                                                        \ \big] \nonumber\\ 
    +&\ s^4 \big[\ ( 2016 - 5664 d + 1440 d^2 - 96 d^3 ) T 
                + ( 30720 - 13896 d + 1704 d^2 - 48 d^3 ) U^3   \nonumber\\ 
     & \hspace{8mm} 
                + (  - 110016 + 50720 d - 6816 d^2 + 256 d^3 ) U D 
                                                        \ \big] \nonumber\\ 
    +&\ s^3 \big[\ (  - 68224 + 18432 d - 224 d^2 - 96 d^3 ) D^2 
                + (  - 13280 + 3272 d + 146 d^2 - 38 d^3 ) U^4  \nonumber\\ 
     & \hspace{8mm} 
                + ( 70176 - 17696 d - 528 d^2 + 176 d^3 ) U^2 D 
                + (  - 576 - 6720 d + 2880 d^2 - 192 d^3 ) U T 
                                                        \ \big] \nonumber\\ 
    +&\ s^2 \big[\ (  - 24960 - 74112 d + 23424 d^2 - 1152 d^3 ) D T 
                + ( 624 + 828 d - 228 d^2 ) U^5                 \nonumber\\ 
     & \hspace{8mm} 
                + ( - 4992 - 6624 d + 1824 d^2 ) U^3 D 
                + ( 6240 + 18528 d - 5856 d^2 + 288 d^3 ) U^2 T \nonumber\\ 
     & \hspace{8mm} 
                + ( 9984 + 13248 d - 3648 d^2 ) U D^2   \ \big] \nonumber\\ 
    +&\ s\ 
           \big[\ (  - 9984 d + 7296 d^2 - 1152 d^3 ) D^3 
                + ( 156 d - 114 d^2 + 18 d^3 ) U^6              \nonumber\\ 
     & \hspace{8mm} 
                + (  - 1872 d + 1368 d^2 - 216 d^3 ) U^4 D 
                + ( 7488 d - 5472 d^2 + 864 d^3 ) U^2 D^2   \ \big] 
\ , \labbel{eq:defC2} 
\end{align} 
\begin{align}
 C_1(d,s,m_i^2) &= 
      s^6 ( 9408 - 6440 d + 1610 d^2 - 175 d^3 + 7 d^4 )        \nonumber\\ 
   &+ s^5 ( 1440 - 3792 d + 1708 d^2 - 270 d^3 + 14 d^4 ) U     \nonumber\\ 
   &+ s^4 \big[\ ( 104256 - 77088 d + 20536 d^2 - 2316 d^3 + 92 d^4 ) D 
                                                                \nonumber\\ 
   &\hspace{8mm}+( - 36480 + 27352 d - 7182 d^2 + 767 d^3 - 27 d^4 ) U^2 
                                                          \big] \nonumber\\ 
   &+ s^3 \big[\ ( - 4032 + 4384 d - 1888 d^2 + 416 d^3 - 32 d^4 ) T 
                                                                \nonumber\\ 
   &\hspace{8mm}+( 29760 - 16688 d + 2712 d^2 - 52 d^3 - 12 d^4 ) U^3 
                                                                \nonumber\\ 
   &\hspace{8mm}+( - 117888 + 69440 d - 13024 d^2 + 688 d^3 + 16 d^4 ) U D 
                                                          \big] \nonumber\\ 
   &+ s^2 \big[\ ( - 29952 + 10368 d + 736 d^2 - 720 d^3 + 80 d^4 ) D^2 
                                                                \nonumber\\ 
   &\hspace{8mm}+( - 2880 - 1416 d + 1342 d^2 - 273 d^3 + 17 d^4 ) U^4 
                                                                \nonumber\\ 
   &\hspace{8mm}+( 19008 + 3072 d - 5552 d^2 + 1272 d^3 - 88 d^4 ) U^2 D 
                                                                \nonumber\\ 
   &\hspace{8mm}+( 1152 + 576 d - 1472 d^2 + 576 d^3 - 64 d^4 ) U T 
                                                          \big] \nonumber\\ 
   &+ s \ \big[\ ( 49920 - 23168 d - 7296 d^2 + 3968 d^3 - 384 d^4 ) D T 
                                                                \nonumber\\ 
   &\hspace{8mm}+( - 1248 + 1088 d - 292 d^2 + 34 d^3 - 2 d^4 ) U^5 
                                                                \nonumber\\ 
   &\hspace{8mm}+( 9984 - 8704 d + 2336 d^2 - 272 d^3 + 16 d^4 ) U^3 D 
                                                                \nonumber\\ 
   &\hspace{8mm}+( - 12480 + 5792 d + 1824 d^2 - 992 d^3 + 96 d^4 ) U^2 T 
                                                                \nonumber\\ 
   &\hspace{8mm}+( - 19968 + 17408 d - 4672 d^2 + 544 d^3 - 32 d^4 ) U D^2 
                                                          \big] \nonumber\\ 
   &+\hspace{3mm} \big[\ 
                 ( 6656 d - 6528 d^2 + 1984 d^3 - 192 d^4 ) D^3 \nonumber\\ 
   &\hspace{8mm}+( - 104 d + 102 d^2 - 31 d^3 + 3 d^4 ) U^6     \nonumber\\ 
   &\hspace{8mm}+(  1248 d - 1224 d^2 + 372 d^3 - 36 d^4 ) U^4 D \nonumber\\ 
   &\hspace{8mm}+( - 4992 d + 4896 d^2 - 1488 d^3 + 144 d^4 ) U^2 D^2 \big] 
\ , \labbel{eq:defC1} 
\end{align} 
\begin{align}
 C_0(d,s,m_i^2) &= (d-3)(d-4) \times \big\{                      \nonumber\\ 
   &+ s^5  ( 336 - 146 d + 21 d^2 - d^3 ) 
    + s^4  ( 360 - 258 d + 51 d^2 - 3 d^3 ) U                     \nonumber\\ 
   &+ s^3 \big[\ ( 4752 - 2280 d + 344 d^2 - 16 d^3 ) D 
              + ( - 1280 + 568 d - 70 d^2 + 2 d^3 ) U^2\ \big]    \nonumber\\ 
   &+ s^2 \big[\ ( 336 - 104 d - 48 d^2 + 8 d^3 ) T 
              + ( 240 + 120 d - 66 d^2 + 6 d^3 ) U^3              \nonumber\\ 
   &\hspace{8mm}+( - 1056 - 400 d + 248 d^2 - 24 d^3 ) U D\ \big] \nonumber\\ 
   &+ s\  \big[\ ( 2496 - 2240 d + 592 d^2 - 48 d^3 ) D^2 
              + ( 240 - 182 d + 33 d^2 - d^3 ) U^4                \nonumber\\ 
   &\hspace{8mm}+(  - 1584 + 1288 d - 280 d^2 + 16 d^3 ) U^2 D 
              + (  - 96 + 176 d - 96 d^2 + 16 d^3 ) U T \   \big] \nonumber\\ 
   &+\hspace{3mm} \big[\ 
                (  - 4160 + 3872 d - 1088 d^2 + 96 d^3 ) D T 
              + ( 104 - 102 d + 31 d^2 - 3 d^3 ) U^5              \nonumber\\ 
   &\hspace{8mm}+( - 832 + 816 d - 248 d^2 + 24 d^3 ) U^3 D 
              + ( 1040 - 968 d + 272 d^2 - 24 d^3 ) U^2 T         \nonumber\\ 
   &\hspace{8mm}+( 1664 - 1632 d + 496 d^2 - 48 d^3 ) U D^2 \ \big] 
 \big\} \ , \labbel{eq:defC0} 
\end{align} 
\par 
For completeness, we report here the explitcit expressions of some 
coefficients appearing in Section \ref{sec:DD2}, which are also 
symmetrical in the three masses. 
According to Eq.(\ref{eq:D42}) of Section \ref{sec:DD2}, at $ d=2 $ the 4th 
order operator $ D_4(d,s,m_i^2) $, Eq.(\ref{eq:D4}) or Eq.(\ref{eq:D4a}) 
can be factorised as 
\be D_4(2,s,m_i^2) =   \mathcal{K}\ D_{1b}(2,s,m_i^2) 
                     \ D_{1a}(2,s,m_i^2)\ D_2(2,s,m_i^2)\ , \labbel{eq:D42a} 
\ee 
with 
\be D_2(2,s,m_i^2) = B_2(2,s,m_i^2) \frac{d^2}{ds^2} 
                   + B_1(2,s,m_i^2) \frac{d}{ds} 
                   + B_0(2,s,m_i^2) \ . \labbel{eq:D22} 
\ee 
In the notation of the previous formulas, the explicit expressions 
of the coefficients $ B_k(2,s,m_i^2) $ are 
\begin{align} 
 B_2(2,s,m_i^2) &= s\ \mathcal{D}(s)\ ( -3s^2 + 2s U +  U^2-4D ) \ , 
                                                                  \nonumber\\ 
 B_1(2,s,m_i^2) &= -9s^6 + 32s^5 U + s^4( -37 U^2 + 4D ) 
                                   + s^3( 32 UD + 8U^3 )          \nonumber\\ 
                &+ s^2( - 128UT - 88U^2D + 13U^4 + 144D^2 )       \nonumber\\ 
                &+ s  ( - 128UD^2 - 128U^2T + 64U^3D - 8U^5 + 512DT ) 
                                                                  \nonumber\\ 
                & + 48U^2D^2 - 12U^4D + U^6 - 64D^3 \ ,           \nonumber\\ 
 B_0(2,s,m_i^2) &= -2s^5 + 7s^4 U + s^3( -2U^2 - 12D ) 
                                  + s^2( 32UD - 6U^3 - 60T )      \nonumber\\ 
                &+ s^2 ( 32UD - 6U^3 - 60T ) 
                 + s   ( 8UT - 28U^2D + 5U^4 + 32D^2 )            \nonumber\\ 
                &- 16UD^2 - 12U^2T + 8U^3D - U^5 + 48DT \ . \labbel{eq:B2i} 
\end{align} 
\section{The $b$ integrals at $ d=2 $ as elliptic functions} \labbel{app:d2} 
\setcounter{equation}{0} 
\numberwithin{equation}{section} 
According to the definitions Eq.s(\ref{eq:Ni},\ref{eq:R4}), at $ d= 2 $ 
the six $b$ integrals become 
\be N_i(2,s) = \int_{b_{i-1}}^{b_i} \frac{db} 
        {\sqrt{|(b-b_1)(b-b_2)(b-b_3)(b-b_4)|}} , \hspace{5mm} i=0,1,..,5 
\labbel{eq:d2a} \ee 
where the $b_i$'s, given by Eq.s(\ref{eq:bi}), are in increasing order 
\be b_{-1} < b_0 < b_1 < b_2 < b_3 < b_4 < b_5 \ . \labbel{eq:d2biord} \ee 
The six integrals can be expressed, see~\cite{Caleca2020} 
by using the Legendre changes of 
integration variable, in terms of elliptic integrals of the first kind, 
defined as
\begin{equation}
    \begin{split}
        F(x,k) &= \int_0^x \frac{dt}{\sqrt{(1-t^2)(1-k^2t^2)}} \ , \\
        K(k) &= \int_0^1 \frac{dt}{\sqrt{(1-t^2)(1-k^2t^2)}} \ , 
                \labbel{eq:d2FK}
    \end{split} 
\end{equation} 
where $ F(x,k) $ is the incomplete elliptic integral (with phase $x$) 
and $ K(k) $ is the complete elliptic integral. $ K(k) $ is a solution of the 
2nd order differential equation 
\be \left[ k(1-k^2) \frac{d^2}{dk^2} 
           + (1-3k^2)\frac{d}{dk} - k \right] K(k) = 0 \ ,  
    \labbel{eq:d2equK} 
\ee 
{\it i.e.} 
\be k(1-k^2) K''(k) + (1-3k^2) K'(k)- k K(k) = 0 \ ,  
    \labbel{eq:d2equKk} 
\ee 
with a second solution of the same equation provided by $ K(k') $, with 
\be k'^2 = 1 - k^2 \ . \labbel{eq:d2k'} \ee 
\par 
For $N_0(2,s)$ and $N_1(2,s)$, the change of integration variable is 
\begin{equation} 
    t^2 = \frac{(b_4-b_2)(b_1-b)}{(b_4-b_1)(b_2-b)} \ , \hspace{5mm} 
      b = \frac{(b_4-b_2)b_1-(b_4-b_1)b_2t^2}
               {(b_4-b_2)   -(b_4-b_1)t^2} \ , \labbel{eq:d2t01} 
\end{equation} 
giving 
\begin{align}
     & N_0(2,s) = \frac{2}{\sqrt{(b_4-b_2)(b_3-b_1)}} \Bigg[
    F\bigg(\sqrt{\frac{b_4-b_2}{b_4-b_1}},q\bigg) -
    F\bigg(\sqrt{\frac{b_1(b_4-b_2)}{b_2(b_4-b_1)}},q\bigg) \bigg] \ , \\ 
    & N_1(2,s) = \frac{2}{\sqrt{(b_4-b_2)(b_3-b_1)}} \ 
    F\bigg(\sqrt{\frac{b_1(b_4-b_2)}{b_2(b_4-b_1)}},q\bigg) \ , 
\labbel{eq:d201} \end{align}
with 
\begin{equation}
    q^2 = \frac{(b_4-b_1)(b_3-b_2)}{(b_3-b_1)(b_4-b_2)}. \labbel{eq:d2q} 
\end{equation} 
\par 
For $N_2(2,s)$ the change of variable is
\begin{equation}
    t^2 = \frac{(b_4-b_2)(b-b_1)}{(b_2-b_1)(b_4-b)} \ , \labbel{eq:d2t2} 
\end{equation} 
giving 
\begin{equation} 
    N_2(2,s) = \frac{2}{\sqrt{(b_4-b_2)(b_3-b_1)}} \ K(q') \ , \labbel{eq:d22} 
\end{equation} 
where 
\begin{equation}
    q'^2 = 1 - q^2 = \frac{(b_4-b_3)(b_2-b_1)}{(b_3-b_1)(b_4-b_2)} \ . 
\labbel{eq:d2qq} \end{equation} 
\par 
For $N_3(2,s)$ one proceeds in a similar way. By writing 
\begin{equation} 
    t^2 = \frac{(b_3-b_1)(b-b_2)}{(b_3-b_2)(b-b_1)} \ , \labbel{eq:d2t3} 
\end{equation}
the result is 
\begin{equation} 
    N_3(2,s) = \frac{2}{\sqrt{(b_4-b_2)(b_3-b_1)}}  \ K(q)\ . \labbel{eq:d23} 
\end{equation} 
\par 
For $N_4(2,s)$ the change of variable is 
\begin{equation}
    t^2 = \frac{(b_3-b_1)(b_4-b)}{(b_4-b_3)(b-b_1)} \ , \labbel{eq:d2t4} 
\end{equation} 
which gives 
\begin{equation} 
    N_4(2,s) = \frac{2}{\sqrt{(b_4-b_2)(b_3-b_1)}} \ K(q')\ , \labbel{eq:d24} 
\end{equation} 
equal to $ N_3(2,s), $ according to Eq.(\ref{eq:M2M4}). 
\par 
Finally, for $N_5(2,s)$ the change of variable is 
\begin{equation}
    t^2 = \frac{(b_3-b_1)(b-b_4)}{(b_4-b_1)(b-b_3)} \ , \labbel{eq:d2t5} 
\end{equation} 
and the result is 
\begin{equation} 
    N_5(2,s) = \frac{2}{\sqrt{(b_4-b_2)(b_3-b_1)}} \ 
    F\bigg(\sqrt{\frac{b_3-b_1}{b_4-b_1}},q\bigg)\ . \labbel{eq:d25} 
\end{equation} 
\par 
Let us observe that Eq.s(\ref{eq:d22},\ref{eq:d23},\ref{eq:d24}) are of 
the form 
\be N(s) = f(s) K(g(s)) \ , \labbel{eq:NtoK} \ee 
where $ f(s), g(s) $ are known and simple functions of $ s $ (involving 
at most square roots). By using the chain differentiation rule, one can 
therefore differentiate repeatedly Eq.(\ref{eq:NtoK}) with respect to $s$, 
and express in that way the derivatives of $ K(g(s)) $ in terms of the 
derivatives of $ N(s) $ times known coefficients. By inserting those 
expressions into Eq.(\ref{eq:d2equKk}) one obtains a second order 
differential equation in $ s $ for $ N(s) $, {\it i.e.} for 
$ N_2(2,s) = N_4(2,s) $ and $ N_3(2,s) $. That equation coincides (of course) 
with Eq.(\ref{eq:D22}) or, which is the same, Eq.(\ref{eq:D2M3}) of 
Section \ref{sec:DD2}. 
\par 
As a further remark, Eq.(\ref{eq:M0135}) can be written as 
\be N_0(2,s) + N_1(2,s) + N_5(2,s) = N_3(2,s) \ , \labbel{eq:d2N0153} \ee 
which in terms of the above elliptic functions reads 
\be F\bigg(\sqrt{\frac{b_4-b_2}{b_4-b_1}},q\bigg) 
  + F\bigg(\sqrt{\frac{b_3-b_1}{b_4-b_1}},q\bigg) = K(q) \ . 
\labbel{eq:d2N0153a} \ee 
\bibliographystyle{bibliostyle}
\bibliography{Biblio} 
\end{document}